\documentclass[aip,jcp,amsmath,amssymb,reprint]{revtex4-1}

\usepackage{graphicx}
\usepackage[separate-uncertainty=true]{siunitx}
\graphicspath{{figures/}}

\begin{document}

\title{Decoding the spectroscopic features and timescales of aqueous proton defects}

\author{Joseph A. Napoli}
\affiliation{Department of Chemistry, Stanford University, Stanford, California 94305, USA}

\author{Ondrej Marsalek}
\affiliation{Department of Chemistry, Stanford University, Stanford, California 94305, USA}

\author{Thomas E. Markland}
\email{tmarkland@stanford.edu}
\affiliation{Department of Chemistry, Stanford University, Stanford, California 94305, USA}

\date{\today}

\begin{abstract}
Acid solutions exhibit a variety of complex structural and dynamical features arising from the presence of multiple interacting reactive proton defects and counterions. However, disentangling the transient structural motifs of proton defects in the water hydrogen bond network and the mechanisms for their interconversion remains a formidable challenge. Here, we use simulations treating the quantum nature of both the electrons and nuclei to show how the experimentally observed spectroscopic features and relaxation timescales can be elucidated using a physically transparent coordinate that encodes the overall asymmetry of the solvation environment of the proton defect. We demonstrate that this coordinate can be used both to discriminate the extremities of the features observed in the linear vibrational spectrum and to explain the molecular motions that give rise to the interconversion timescales observed in recent nonlinear experiments. This analysis provides a unified condensed-phase picture of proton structure and dynamics that, at its extrema, encompasses proton sharing and spectroscopic features resembling the limiting Eigen [H$_{3}$O(H$_{2}$O)$_{3}$]$^{+}$ and Zundel [H(H$_{2}$O)$_{2}$]$^{+}$ gas-phase structures, while also describing the rich variety of interconverting environments in the liquid phase.
\end{abstract}

\maketitle

The structure and transport of proton defects in aqueous solution underpin processes ranging from voltage-gated channels in biology to the development of improved proton exchange membrane materials~\cite{Kreuer1996/10.1021/cm950192a,Decoursey2003/10.1152/physrev.00028.2002,Wraight2006/10.1016/j.bbabio.2006.06.017,Peighambardoust2010/10.1016/j.ijhydene.2010.05.017}. While it has long been appreciated that proton defects form a wide range of structures whose interconversion enables rapid proton transport, identification of the dominant motifs and their timescales remains a significant experimental~\cite{Headrick2004/10.1063/1.1834566,Headrick2005/10.1126/science.1113094,Woutersen2006/10.1103/PhysRevLett.96.138305,Thamer2015/10.1126/science.aab3908,Dahms2016/10.1002/anie.201602523,Agmon2016/10.1021/acs.chemrev.5b00736} and theoretical~\cite{Tuckerman1995/10.1021/j100016a003,Marx1999/10.1038/17579,Iftimie2006/10.1002/anie.200502259,Kaledin2006/10.1021/jp054374w,Chandra2007/10.1103/PhysRevLett.99.145901,Xu2010/10.1021/jp102516h,Xu2011/10.1021/jz101536b,Kale2012/10.1002/anie.201203568,Hassanali2013/10.1073/pnas.1306642110,Kulig2013/10.1038/nchem.1503,Baer2014/10.1021/jp501091h,Baer2014/10.1021/jp501854h,Mancini2015/10.1039/C4CP05685J,Biswas2017/10.1063/1.4980121,Agmon2016/10.1021/acs.chemrev.5b00736} challenge. 
Linear vibrational spectroscopies provide one approach to probe these features. However, the electronic and structural flexibility of proton defects gives rise to a broad and largely featureless continuum between the far infra-red (IR) and the O-H stretch band regions of the absorption spectrum (see Fig.~\ref{fig:linear_comp}a). 
Indeed, the flexibility of proton defects has been highlighted in gas-phase studies where even subtle chemical perturbations around the defect have been observed to induce significant spectral changes~\cite{Headrick2005/10.1126/science.1113094,Olesen2011/10.1016/j.cplett.2011.04.060,Craig2017/10.1073/pnas.1705089114}. This raises the question as to whether the broad linear vibrational spectrum in solution arises simply from rapid fluctuations of the proton defect structure or whether multiple defect structures exist that undergo slower interconversion with each contributing their own spectral signatures.

Separating these two pictures requires knowledge of the timescales for the interconversion of proton defect structures. Nonlinear spectroscopy provides a route to obtain this more detailed information by allowing measurement of the correlation times of the frequencies associated with proton defects. Specifically, recent two-dimensional (2D) IR experiments on hydrochloric (HCl) acid solutions probed a stretch-bend cross-peak and observed a timescale of 480~fs, which was interpreted as a lower bound on the Zundel complex lifetime~\cite{Thamer2015/10.1126/science.aab3908}. While this result suggests that certain proton defect structures interconvert over timescales longer than previously observed~\cite{Woutersen2006/10.1103/PhysRevLett.96.138305}, the interpretation of the cross-peak decay rested largely on assigning frequencies to the Zundel gas-phase structure within a cluster-centric paradigm~\cite{Kim2002/10.1063/1.1423327,Xu2011/10.1021/jz101536b,Biswas2017/10.1063/1.4980121,Biswas2016/10.1063/1.4964723,Lin2015/10.1021/jp509791n,Kulig2013/10.1038/nchem.1503}. However, it remains unclear whether an interpretation of the liquid spectra based on the limiting Eigen and Zundel gas-phase cluster structures provides a complete picture of the structural motifs of condensed-phase proton defects and their interconversion timescales.

Here we use classical and path integral ab initio molecular dynamics simulations to decode the linear and nonlinear vibrational spectra of concentrated aqueous acid solutions. In particular, we show that one can define a simple coordinate that reveals how the vibrational spectrum and its evolution timescales arise from the asymmetry of the hydration structure of proton defects.

\section{Results}

\begin{figure*}[ht]
  \centering
  \includegraphics{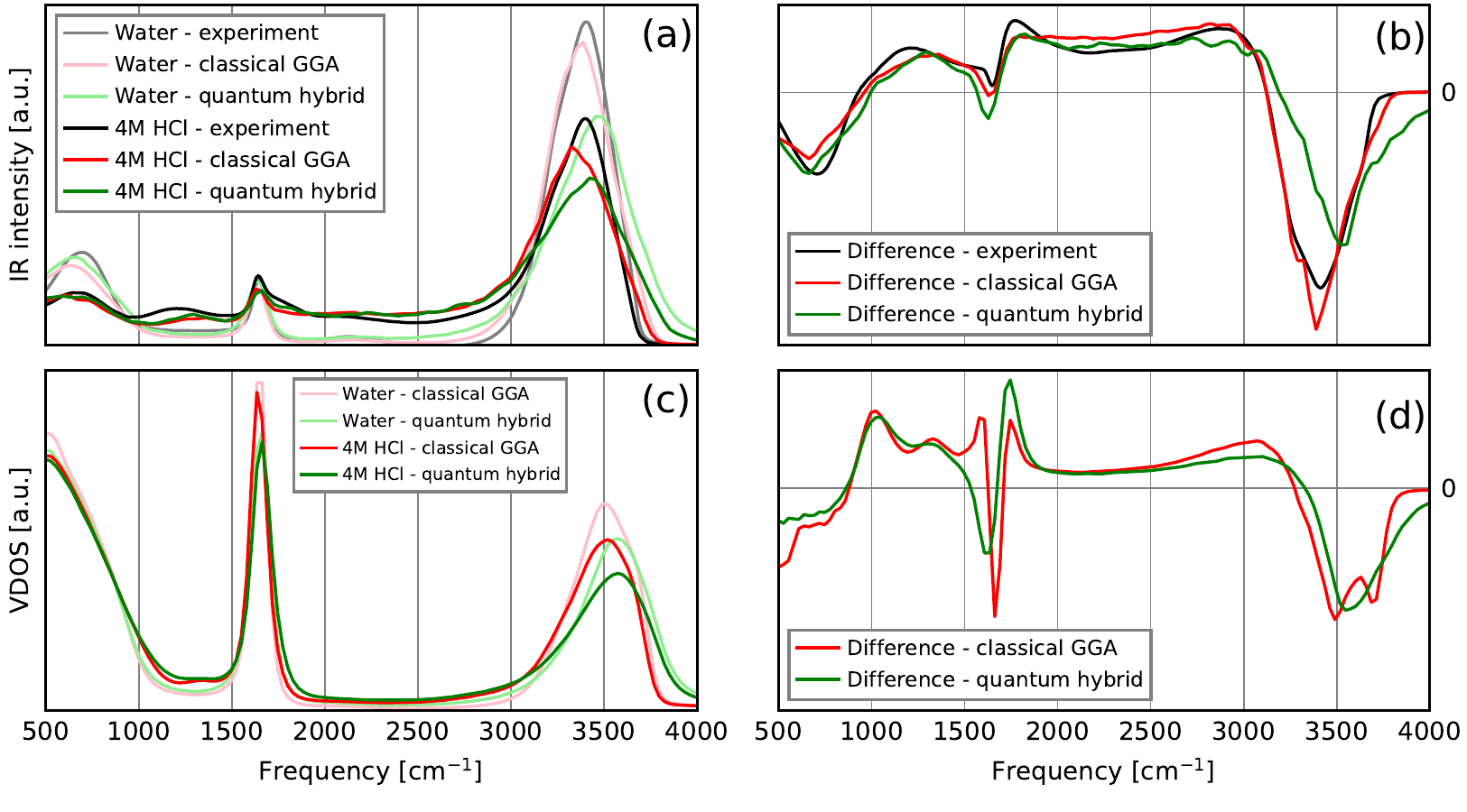}
  \caption{\label{fig:linear_comp}
    IR and VDOS spectra and difference spectra for 4M HCl solution and neat water from experiment and ab initio molecular dynamics simulations with quantum and classical nuclei. (a) Experimental and simulated linear IR spectra with each normalized to unit area over the range shown. (b) Experimental and simulated difference IR spectra obtained from the difference of the normalized acid and water spectra. (c) Vibrational density of states (VDOS) spectra. (d) Difference VDOS spectra obtained from the normalized VDOS. Quantum indicates ab initio molecular dynamics simulations in which the nuclear dynamics were treated quantum mechanically using TRPMD. Hybrid and GGA denote the level of the electronic structure approach employed.}
\end{figure*}

We first consider the experimental and simulated linear IR spectra of 4M HCl solution and neat water in Fig.~\ref{fig:linear_comp}a. The spectra were calculated directly from the dipole correlation function obtained from ab initio molecular dynamics simulations. The quantum results were obtained from TRPMD simulations, which include quantum effects of the nuclei such as zero-point energy and tunnelling, while the classical ones do not. While the classical simulations employed the GGA revPBE-D3 functional, due to the large zero-point energy contribution that is introduced when the nuclei are treated quantum mechanically ($\sim$5 kcal/mol per hydrogen bond), the more accurate hybrid revPBE0-D3 functional was employed for the TRPMD simulation to avoid spurious effects that are encountered when quantizing GGA functionals~\cite{Marsalek2017/10.1021/acs.jpclett.7b00391}.  

Both the simulated and experimental spectra exhibit continuous absorption between the water bend at $\sim$1600~cm$^{-1}$ and the beginning of the water O--H stretch at $\sim$3000~cm$^{-1}$. Taking the difference between the acid and water spectra allows one to identify distinct proton spectral features (Fig.~\ref{fig:linear_comp}b). In particular, in addition to the increased intensity between the water bend and stretch, the bend broadens and a feature emerges around 1250~cm$^{-1}$. This feature, which has previously been suggested to be a Zundel-like shuttling motion~\cite{Asmis2003/10.1126/science.1081634,Headrick2004/10.1063/1.1834566,Headrick2005/10.1126/science.1113094}, corresponds to proton motion orthogonal to the O--O axis in our simulations (see Fig.~S1) and is thus associated with a bending motion of the proton defect, consistent with earlier simulation studies~\cite{Biswas2017/10.1063/1.4980121,Iftimie2006/10.1002/anie.200502259}. The ab initio simulations and experiment exhibit excellent agreement for both the total and difference spectra over the entire region for which the experimental data are available ($>$600~cm$^{-1}$), with the small discrepancy in the intensity depletion around 2250~cm$^{-1}$ arising from the underestimation of the bend-libration coupling feature in the simulated water spectrum. The features of the IR spectrum, computed from the dipole autocorrelation function~\cite{McQuarrie}, arise from the underlying motions of the nuclei and are thus also encoded in the vibrational density of states (VDOS) of the system. This is shown in panels (c) and (d) of Fig.~\ref{fig:linear_comp}, where the absolute and difference VDOS spectra exhibit the same structural features as the IR, albeit with different intensities.

\begin{figure}[ht]
  \centering
  \includegraphics{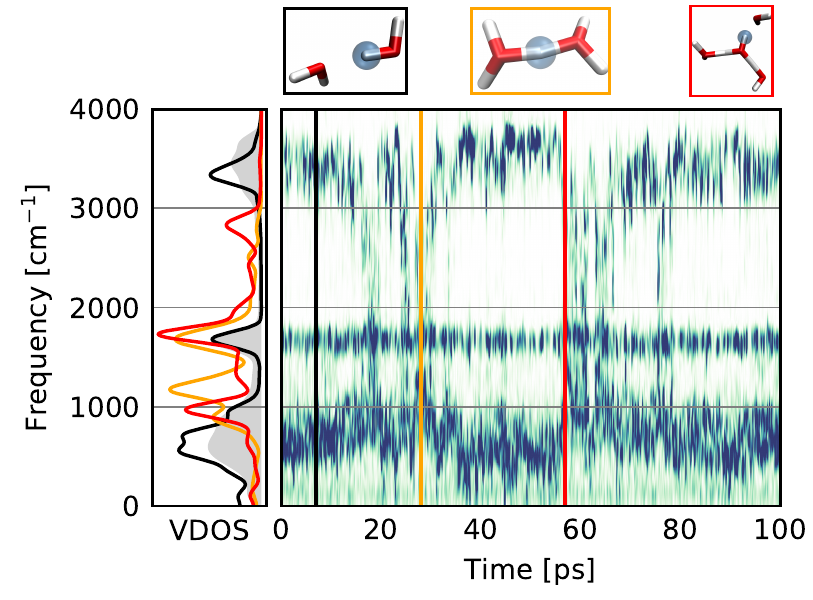}
  \caption{\label{fig:VDOS_t}
    Time evolution of the spectrum of a single proton in a 4M HCl solution over a 100~ps segment of the simulation. When the proton participates in a defect structure proton spectral features are observed. The vertical black (7~ps), orange (28~ps), and red (57~ps) lines in the main panel correspond to the left, middle, and right simulation snapshots at the top, respectively, which show the local proton environment in solution at that time. The corresponding normalized spectra of the proton at the points in time corresponding to the vertical lines are shown in the left panel, with the total VDOS of that proton over the entire trajectory shown as gray shading for reference.
  }
\end{figure}

Given the correspondence between the VDOS and IR spectra, we therefore consider the time evolution of the VDOS of the protons in the acid solution to assess how proton spectral features arise and interconvert. Figure~\ref{fig:VDOS_t} shows the time evolution of the spectrum of one proton in a 4M HCl solution over a 100~ps simulation segment. During this segment, the proton is initially part of a water molecule (black), is part of a proton defect structure at $\sim$28~ps (orange), after which it returns to being part of a water molecule, and finally returns to being part of a proton defect (red). It then continues to fluctuate in and out of proton defect environments throughout the remainder of the simulation segment. The left hand panel shows the spectrum associated with the proton at the three times shown (black, orange, red) compared to the total spectrum that proton experiences over the 100~ps simulation segment (grey). From this, one can see that the proton spectral features identified in Fig.~\ref{fig:linear_comp} are observed when the proton forms part of a proton defect structure (orange, red). The time evolution of the VDOS thus provides a way to associate the vibrational spectrum of a specific proton over the entire frequency range and its evolution with the continually shifting local structural features in the condensed-phase environment of the acid solution. 

The time evolution of the simulated spectrum can then be used to compute the frequency-frequency correlation function corresponding to the pump (3150~cm$^{-1}$) and probe (1760~cm$^{-1}$) frequencies used in Ref.~\onlinecite{Thamer2015/10.1126/science.aab3908}. Doing this for our aqueous excess proton defect simulation, one obtains a time constant of 1.4~$\pm$~0.3~ps (see Fig.~S3), which is consistent with the experimentally observed bound of $>$480~fs~\cite{Thamer2015/10.1126/science.aab3908}. The self-correlation time of the probe frequency (1760~cm$^{-1}$) obtained from our simulation is 1.6~$\pm$~0.3~ps (see Fig.~S4), suggesting that, consistent with the experimental interpretation, the primary mechanism for the slow relaxation timescale of that feature is via interconversion to a species that absorbs strongly at 3150~cm$^{-1}$. 

\begin{figure}[ht]
  \centering
  \includegraphics{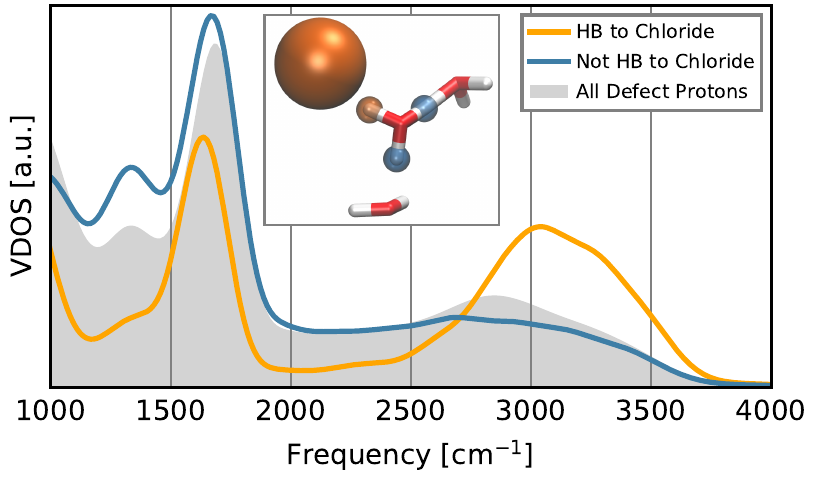}
  \caption{\label{fig:ion_pair}
    Spectra of proton defects adjacent to chloride. The VDOS of protons that are hydrogen bonded to chloride is shown in orange, the VDOS of protons not hydrogen bonded to chloride is shown in blue, and the VDOS of all protons participating in any defect in the 4M HCl solution is shown as gray shading for reference. The inset shows a representative simulation snapshot of a proton defect hydrogen bonded to a chloride anion, with the hydrogen atoms highlighted in the colors corresponding to the curves.
  }
\end{figure}

What is the molecular origin of this timescale and what structural interconversion is it probing? To assess this, we consider the subensemble of protons directly connected to overcoordinated oxygen atoms (see Sec.~\ref{sec:Methods}). There are three such protons by construction, which we refer to as the defect protons. While this allows us to define a subensemble of defect protons, it does not bias our subsequent analysis to any particular structure of the proton defect. Having done this, it is instructive to first consider the spectra of protons in defects where the hydrogen bond environment around the defect is explicitly asymmetric due to the formation of a hydrogen bond to a chloride ion (Fig.~\ref{fig:ion_pair}). Comparing the spectra of protons which are hydrogen bonded to the chloride (orange) and those that are not (blue), we observe that the former exhibit reduced, while the latter exhibit enhanced, proton spectral features, compared to the whole proton defect subensemble. This reflects the greater ability of the water oxygen atoms to share the defect protons, compared to the chloride ion, and thus underscores the impact of environment asymmetries on the spectral features of the proton defect. However, while the presence of a chloride ion in the proton defect's 1$^{\text{st}}$ coordination shell explicitly breaks the symmetry of its chemical environment, the case of a defect coordinated solely by water is more subtle.

\begin{figure}[ht]
  \centering
  \includegraphics[scale=0.5]{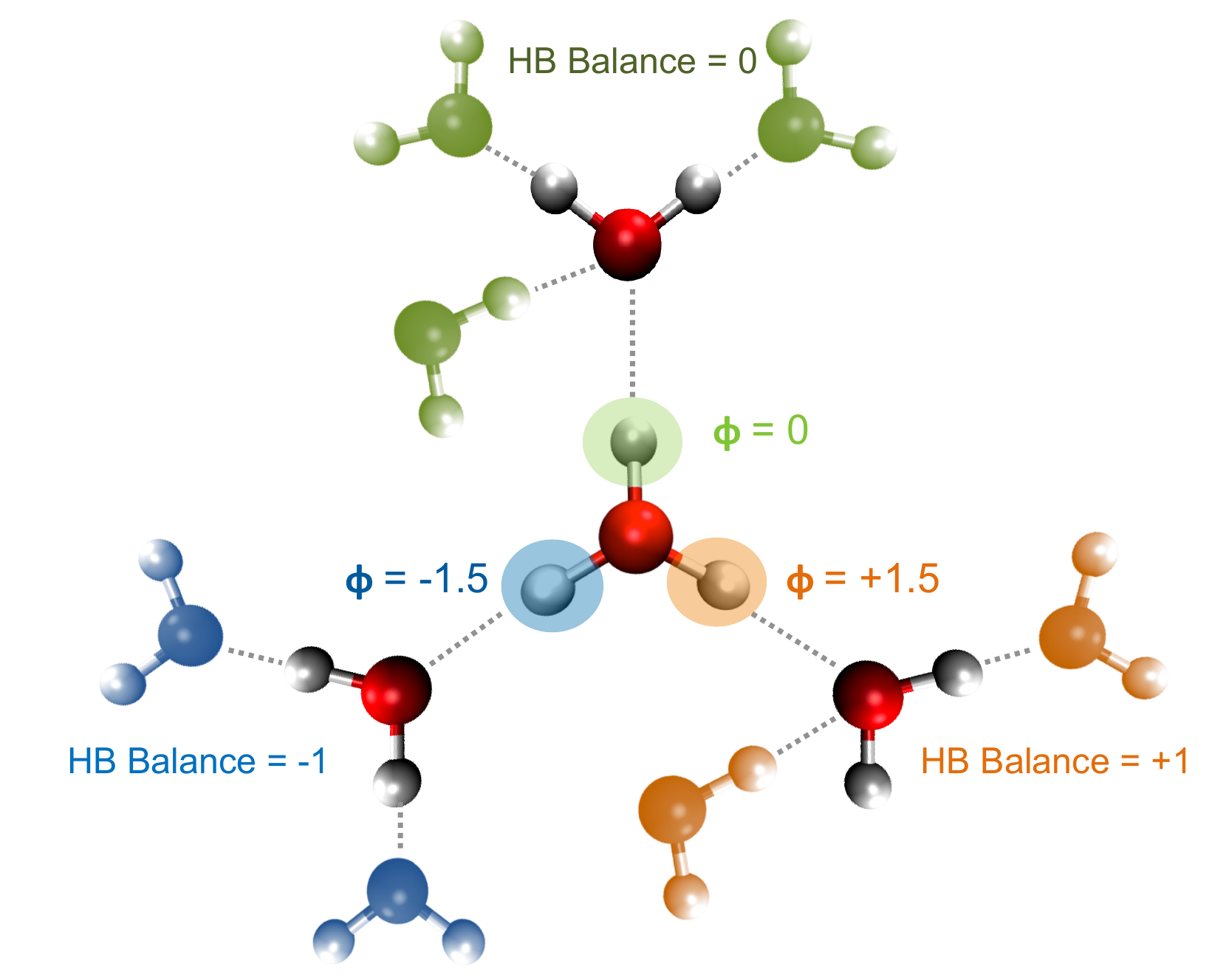}
  \caption{\label{fig:phi_example}
    Illustration explaining the definition of the asymmetry coordinate $\phi$.
    The protons connected to the overcoordinated oxygen form hydrogen bonds to three waters. Each water molecule in the 1$^{\text{st}}$ coordination shell of the defect is assigned a hydrogen bond balance, defined as the difference between the number of hydrogen bonds it accepts and the number it donates. For example, the water molecule surrounded by green shaded waters at the top accepts one hydrogen bond from a defect proton, accepts another from a water molecule, and donates two hydrogen bonds. Its hydrogen bond balance is thus 0. Each defect proton is shaded in a color matching the environment it points at. For example for the green defect proton, $\phi = 0 - (-1 + 1) / 2 = 0$. Each defect proton is assigned a $\phi$ value, and by construction the sum of those values equals 0.     
  }
\end{figure}

We now consider how broken symmetry around proton defects coordinated solely by water molecules manifests spectroscopically. In this case, all three acceptors are chemically identical in the 1$^{\text{st}}$ coordination shell of the defect (Fig.~\ref{fig:phi_example}). One approach to differentiating the defect protons' solvation environments is to consider the ``hydrogen bond balance'' of the water molecules they point at. In particular, the hydrogen bond balance is defined as the difference between the number of hydrogen bonds a water molecule receives and the number it donates (Fig.~\ref{fig:phi_example}). This definition, which measures whether a water molecule is under or overcoordinated and hence indicates its propensity to accept or donate additional hydrogen bonds, has been successfully used to analyze the properties of numerous hydrogen bonded systems~\cite{Lapid2005/10.1063/1.1814973,Ohno2005/10.1039/b506641g,Sun2009/10.1016/j.vibspec.2009.05.002,Vacha2012/10.1021/jz2014852,Tainter2013/10.1021/jz301780k}. An illustration of this procedure is shown in Fig.~\ref{fig:phi_example} for three defect protons (blue, green, orange), each of which points at a water with a different hydrogen bond balance arising from its (correspondingly colored) solvation environment. However, considering only the hydrogen bond balances of the acceptor water molecules at which defect protons point leads to poor separation of the spectrum. This is shown in Fig.~S5, where the spectrum resolved only by the hydrogen bond balance gives near identical spectra in each case. The failure of this property alone to decompose the spectrum can be rationalized by considering cases where, although a defect proton might hydrogen bond to a water molecule with a high or low hydrogen bond balance, the other protons in the defect point at water molecules with the same hydrogen bond balances. One might expect that all these defect protons would exhibit similar sharing since none of them is directed at a water molecule that, based on its hydrogen bond balance, is a better acceptor than the others. The origin of the failure of the hydrogen bond balance to decompose the spectrum thus lies in its neglect of the relative chemical environments of the other defect protons.

Hence, we define a proton asymmetry coordinate, $\phi$, that reports on the asymmetry of the hydrogen bond balances of water molecules in the 1$^{\text{st}}$ coordination shell of the defect. Specifically, for each defect proton, its asymmetry $\phi$ is calculated as the difference between the hydrogen bond balance of the water to which it is hydrogen bonded and the average hydrogen bond balance of the waters to which the other two defect protons are hydrogen bonded. For example, if all three defect protons point at water molecules with the same hydrogen bond balance, then this parameter yields a $\phi$ value of zero for each proton, indicating a lack of asymmetry for all of the protons. The calculation of the proton asymmetry coordinate for the protons in a defect is illustrated in Fig.~\ref{fig:phi_example}. We emphasize that an asymmetry value, $\phi$, is assigned to each defect proton and that the sum of the $\phi$ values for the three defect protons is zero by construction. It is also important that this coordinate does not assume a specific proton defect structure beyond the ability to identify overcoordinated oxygen atoms in the system. Although the illustration in Fig.~\ref{fig:phi_example} could be seen as resembling a solvated Eigen structure, since the $\phi$ parameter is a property of a particular proton and not the oxygen to which it is bound, it continues to be well defined even for protons that transfer. Therefore, for a proton that is highly shared between two oxygen atoms one could equally well represent this figure as the extended Zundel-like depiction in Fig.~S6. In what follows, we demonstrate that this purely structural property allows one to cleanly separate the proton spectral features and elucidate their interconversion timescales.

\begin{figure*}[ht]
  \centering
  \includegraphics{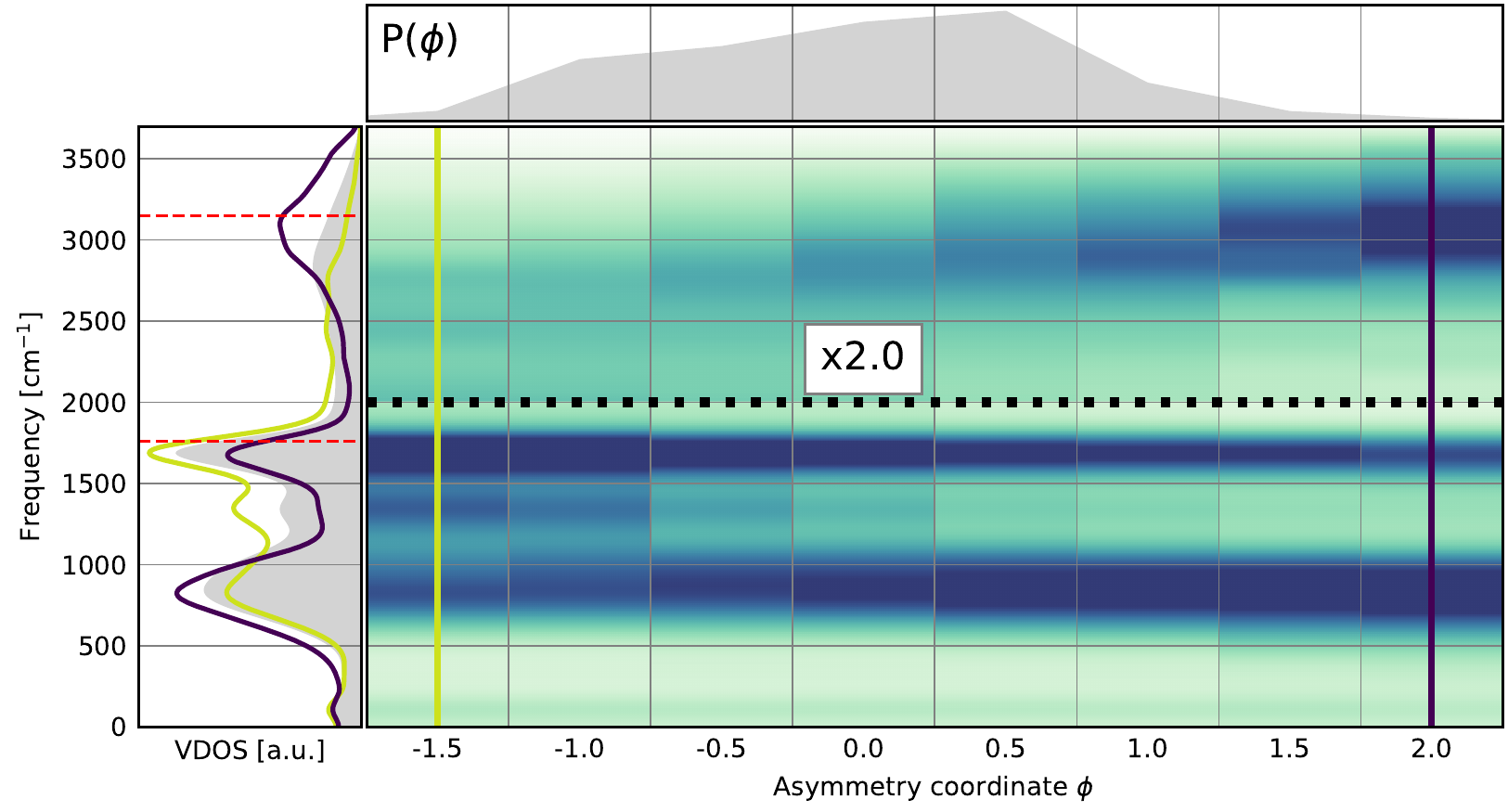}
  \caption{\label{fig:spectrum_phi}
    Decomposition of the defect proton vibrational spectrum (VDOS) with respect to the asymmetry coordinate $\phi$ in a classical ab initio molecular dynamics simulation of an aqueous excess proton. The leftmost bin of the main panel shows the spectrum of protons with very negative values of $\phi$, which exhibits enhanced proton spectral features. The rightmost bin shows that protons with a large positive value of $\phi$ exhibit these considerably less. The spectrum is normalized separately for each value of $\phi$, with the probability distribution of the $\phi$ coordinate displayed in the top panel. The spectra corresponding to the lowest and highest values of $\phi$ are displayed in the left panel in yellow and dark purple, respectively. The spectrum of all defect protons in the simulation is included as gray shading for reference. The region of the main panel corresponding frequencies $>$2000~cm$^{-1}$ is magnified by 2 in order to show the structure in that region.}
\end{figure*}

Figure~\ref{fig:spectrum_phi} shows the defect proton vibrational spectrum resolved by the asymmetry coordinate, $\phi$, for a simulation of an aqueous excess proton defect. The top panel shows the distribution of $\phi$ values obtained from the simulation, which is unimodal and centered around 0, with a range of values from -1.5 to +2.0. The spectrum for the lowest observed value of $\phi$ (yellow line in the left panel) exhibits large enhancements of the 1250~cm$^{-1}$ feature, bend broadening and density in the region between the water bend and stretch, which typify the proton vibrational spectral features observed in Fig.~\ref{fig:linear_comp}. This is because protons with very negative values of $\phi$ point at an acceptor water molecule that donates more hydrogen bonds than it accepts, whereas the other two protons of the defect are directed at water molecules with larger hydrogen bond balances. As such, a proton with a negative value of $\phi$ is hydrogen bonded to an excellent hydrogen bond acceptor water molecule, relative to the acceptors of the hydrogen bonds of the other defect protons. This results in preferential sharing of that proton with its first coordination shell water molecule, which manifests as enhanced proton spectral features. In contrast, the highest observed values of $\phi$ correspond to the opposite case, leading to a low preference for sharing. In this case the spectrum (purple line in the left panel) exhibits few proton spectral features and indeed closely resembles that of the defect protons pointing at a chloride ion in Fig.~\ref{fig:ion_pair}. Hence, consideration of the relative asymmetry of all the hydrogen bonds formed by the defect, as encoded in $\phi$, is essential to understanding the origin of the proton features of vibrational spectra. 

In addition to allowing us to decode the linear spectrum, we can also use the structural parameter $\phi$ to rationalize the observed timescales observed in recent 2D-IR experiments~\cite{Thamer2015/10.1126/science.aab3908}. In particular, one can see from Fig.~\ref{fig:spectrum_phi} that only protons with large values of $\phi$ (purple line) have strong spectral intensity at the pump frequency used in that experiment (3150~cm$^{-1}$), while the intensity is much greater at the probe frequency (1760~cm$^{-1}$) for low $\phi$ protons (yellow line). Hence, the 2D-IR experiment can be interpreted in terms of the $\phi$ coordinate as probing the time it takes for a proton to convert from a high- to low-$\phi$ structure, which is characterized by the $\phi$ correlation time. The $\phi$ correlation time extracted from our simulations (see Fig.~S7) of an aqueous excess proton defect is 1.3~$\pm$~0.3~~ps, in excellent agreement with the frequency-frequency correlation time obtained from our simulations (1.4~ps) and consistent with the 2D-IR experiment ($>$480~fs).

The timescale extracted from the experiment can therefore be understood to arise from hydrogen bond rearrangements in the 1$^{\text{st}}$ and 2$^{\text{nd}}$ hydration shells of the proton defect that decorrelate $\phi$ by changing the overall asymmetry of its environment. The $\phi$ time correlation function, and hence its relaxation timescale, is essentially identical when it is calculated without allowing changes in the overcoordinated oxygen atom (see Fig.~S7). This suggests that the primary mechanism for its decorrelation is water rearrangement around the proton defect rather than proton transfer, which leads to a change in the overcoordinated oxygen~\cite{Chandra2007/10.1103/PhysRevLett.99.145901,Berkelbach2009/10.1103/PhysRevLett.103.238302,Tuckerman2010/10.1063/1.3474625}. This analysis shows that local structural properties of a hydrogen bond (e.g. the hydrogen bond balance) cannot provide a complete picture of the spectroscopy of aqueous protons, which is instead determined by the balance of hydrogen bonds around the defect.

\begin{figure}[ht]
  \centering
  \includegraphics{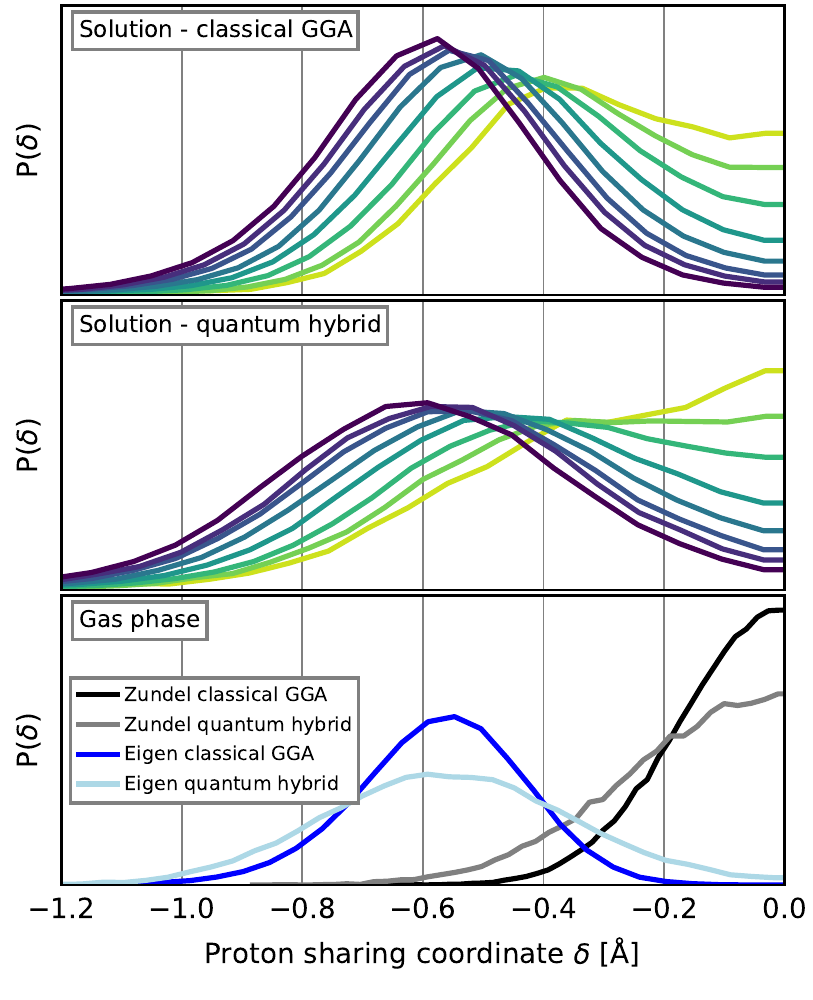}
  \caption{\label{fig:phi_delta}
    Decomposition of the proton sharing coordinate $\delta$ with respect to the asymmetry coordinate $\phi$.
    Distributions of the $\delta$ coordinate of defect protons corresponding to the different $\phi$ values are shown for both classical GGA (top panel) and quantum hybrid (middle panel) simulations of a 4M HCl solution. As $\phi$ progresses from negative (-1.5) to positive (+2.0) in increments of 0.5, the color changes from yellow to dark purple. $\delta$ distributions of classical GGA and quantum hybrid simulations of gas-phase Eigen and Zundel clusters are displayed in the bottom panel for reference.   
  }
\end{figure}

We can now also assess how our approach relates to the commonly invoked paradigm of using the proton sharing coordinate $\delta$, the difference of the distances of the proton from its two nearest oxygen atoms, to classify proton defect structures. Within this approach, structures with highly shared protons ($\delta$ close to zero) are commonly assigned as corresponding to ``Zundel''-type structures, while other structures ($\delta$ further from zero) are assigned to ``Eigen''-type or numerous other purported species~\cite{Xu2011/10.1021/jz101536b,Biswas2017/10.1063/1.4980121,Biswas2016/10.1063/1.4964723,Marx2000/10.1088/0953-8984/12/8A/317,Marx1999/10.1038/17579}. Figure~\ref{fig:phi_delta} shows the distribution of $\delta$ resolved at each $\phi$ value for simulations with classical (top panel) and quantum (middle panel) nuclei. Although values of the proton sharing coordinate $\delta$ close to 0 are more prominent for negative $\phi$ values, there is significant overlap between the distributions of $\delta$, i.e. a given value of $\delta$ could arise when the proton is in many different solvation environments of the proton defect. In each of these environments the proton rapidly explores a large range of $\delta$ values, which leads to the $\delta$ coordinate correlation function for the proton defect decaying to $\sim$0.2 in less than 100~fs (Fig.~S8). This fast decay, corresponding to the exploration of the local environment of a particular proton defect structure, is consistent with that probed in recent ultrafast spectroscopy experiments~\cite{Woutersen2006/10.1103/PhysRevLett.96.138305,Dahms2017/10.1126/science.aan5144}. After this massive and rapid decay, we observe a slower timescale with a time constant of 0.8~$\pm$~0.1~ps, arising from the change in the environmental asymmetry as $\phi$ decorrelates (Fig.~S7). Hence, the decay of the residual of the $\delta$ correlation function does subtly encode a slower timescale consistent with that probed in the 2D-IR experiments, although without providing any insight into the molecular rearrangements that lead to it.

Finally, it is instructive to consider how the distribution along the proton sharing coordinate across the wide range of $\phi$ values in the liquid compares to those of the gas-phase Zundel and Eigen clusters at the same temperature, which are often invoked as the limiting proton defect structures that could exist in aqueous solution. The bottom panel of Fig.~\ref{fig:phi_delta} shows the range of $\delta$ explored by the gas-phase Eigen and Zundel complexes at 300~K with classical and quantum nuclei. Comparing these to the $\phi$-decomposed $\delta$ distributions in the top two panels, we can see that the highest proton asymmetry ($\phi$) values visited in the acid solution have proton sharing ($\delta$) distributions that resemble the Eigen cluster, while the lowest $\phi$ values show substantial population around $\delta = 0$, as for the Zundel cluster. Indeed, it is remarkable that the hydrogen bond asymmetry, which arises from the differing numbers of hydrogen bonds received and donated from water molecules in the 2$^{\text{nd}}$ coordination shell of the proton defect, is enough to give rise to such a broad range of sharing patterns and spectral features of the defect.

\section{Conclusion}
Employing our recent methodological advances~\cite{Marsalek2016/10.1063/1.4941093}, which have enabled us to access nanosecond timescales, we have shown that ab initio molecular dynamics simulations of concentrated acid solutions exhibit excellent agreement with both the experimental IR spectra and 2D-IR relaxation timescales. By decomposing the time-dependent vibrational spectrum of the proton defects across the entire range of frequencies in their full condensed-phase chemical environment, we showed that both the proton spectral features and experimental spectroscopic timescales can be decoded using a structural parameter, $\phi$, which encapsulates the overall solvation asymmetry of a proton's chemical environment. Rather than attempting to rationalize the spectroscopy by considering the properties of individual hydrogen bonds, this parameter provides a collective description of the asymmetry of the proton's solvation, which allows it to distinguish a broad range of proton sharing environments. Perhaps most importantly, it is the extrema of the asymmetry coordinate that give rise to the spectroscopic signatures that coincide with the pump-probe frequency combination studied in recent 2D-IR experiments. The $>$480~fs experimental timescale can therefore be interpreted as emerging from the relaxation of the proton asymmetry coordinate, which occurs by interconversion through a wide range of solvation structures that arise from water hydrogen bond rearrangements in the 2$^{\text{nd}}$ coordination shell. 
This provides a physically transparent framework for decoding the relationship between proton defect structures and their spectroscopic features and timescales in complex chemical environments.

\section{Simulation Details \label{sec:Methods}}

We performed classical and path integral AIMD simulations of concentrated HCl solutions and an aqueous excess proton in the NVT ensemble at T=300~K under periodic boundary conditions using GGA and hybrid density functional theory (DFT) to describe the interactions. Concentrated acid systems of concentrations 2M and 4M were simulated at their experimental densities~\cite{perry-handbook}. For the excess proton system, a cubic box with sides of length 12.42~\AA\ containing 64 water molecules (corresponding to a density of \SI{999.3}{kg.m^{-3}}) and 1 additional proton was simulated, with the initial configurations obtained by protonating equilibrated bulk water. 

Initial structures for the 2M and 4M concentrated acid solutions were obtained from classical molecular dynamics simulations of aqueous chloride ions using the Amoeba2013 polarizable force field~\cite{Shi2013/10.1021/ct4003702}. Two configurations spaced by five nanoseconds were extracted from the trajectories, the required number of water molecules were then protonated and the positions of all protons were minimized. A total of 590~ps of classical simulations were performed for the 2M solution and 420~ps were performed for the 4M solution, composed of trajectories of $\sim$100~ps each. Simulations were performed using the i-PI program~\cite{Ceriotti2013/10.1016/j.cpc.2013.10.027} using a multiple timescale (MTS) integrator of the r-RESPA form~\cite{Tuckerman1992/10.1063/1.463137}. The classical MTS simulations employed a 2.0~fs time step for the full forces and a 0.5~fs time step for the reference forces. Equilibration of 5~ps was performed for each trajectory using a local Langevin thermostat with a time constant of 25~fs, while production runs used a stochastic velocity rescaling thermostat~\cite{Bussi2007/10.1063/1.2408420} with a time constant of 1~ps. 

Path integral AIMD simulations were performed using ring polymer contraction (RPC)~\cite{Markland2008/10.1063/1.2953308,Markland2008/10.1016/j.cplett.2008.09.019,Marsalek2016/10.1063/1.4941093} with centroid contraction, i.e. contraction to $P'=1$ replicas using an MTS propagator with an outer time step of 2.0~fs and an inner time step of 0.25~fs. Simulations performed using contraction to $P'=4$ replicas gave results graphically indistinguishable from those using $P'=1$ for the probability distribution along the proton sharing coordinate (see Fig.~S9). TRPMD path integral simulations~\cite{Craig2004/10.1063/1.1777575,Habershon2013/10.1146/annurev-physchem-040412-110122,Rossi2014/10.1063/1.4883861} were performed by thermostatting the non-centroid normal modes using Langevin thermostats~\cite{Ceriotti2010/10.1063/1.3489925}. A total of over 130~ps of path integral simulations were performed using the revPBE0-D3 functional. The combination of the hybrid revPBE0-D3 functional with path integral simulations was employed due to the spurious effects that are encountered when quantizing GGA functionals such as revPBE-D3~\cite{Marsalek2017/10.1021/acs.jpclett.7b00391}.

The full forces were evaluated using the CP2K program~\cite{Vandevondele2005/10.1016/j.cpc.2004.12.014,Hutter2014/10.1002/wcms.1159} at the DFT level of electronic structure theory using either the revPBE~\cite{Perdew1996/10.1103/PhysRevLett.77.3865,Zhang1998/10.1103/PhysRevLett.80.890} GGA functional or the corresponding revPBE0~\cite{Adamo1999/10.1063/1.478522,Goerigk2011/10.1039/c0cp02984j} hybrid functional, with D3 dispersion corrections~\cite{Grimme2010/10.1063/1.3382344} added in both cases. Atomic cores were represented using the dual-space Goedecker-Tetter-Hutter pseudopotentials~\cite{Goedecker1996/10.1103/PhysRevB.54.1703}. Within the GPW method~\cite{Lippert1997/10.1080/002689797170220}, Kohn-Sham orbitals were expanded in the TZV2P basis set, while an auxiliary plane-wave basis with a cutoff of 400~Ry was used to represent the density. Hybrid functional calculations employed a Coulomb operator truncated~\cite{Guidon2009/10.1021/ct900494g} at $R_{\mathrm{c}}$=6~\AA\ and the auxiliary density matrix method~\cite{Guidon2010/10.1021/ct1002225} with the cpFIT3 fitting basis set. The self-consistent field cycle was converged to an electronic gradient tolerance of $\epsilon_{\mathrm{SCF}} = 5 \times 10^{-7}$ using the orbital transformation method~\cite{VandeVondele2003/10.1063/1.1543154} with the initial guess provided by the always-stable predictor-corrector extrapolation method~\cite{Kolafa2004/10.1002/jcc.10385,Kuhne2007/10.1103/PhysRevLett.98.066401} at each MD step. Full forces for the gas-phase Eigen and Zundel clusters were evaluated using the TeraChem program~\cite{Ufimtsev2009/10.1021/ct9003004} at the DFT level of electronic structure theory using the revPBE-D3~\cite{Perdew1996/10.1103/PhysRevLett.77.3865,Zhang1998/10.1103/PhysRevLett.80.890} GGA functional for classical simulations and the revPBE0-D3~\cite{Adamo1999/10.1063/1.478522,Goerigk2011/10.1039/c0cp02984j} hybrid functional for path integral simulations.

The MTS and RPC reference forces were evaluated at the SCC-DFTB3~\cite{Gaus2011/10.1021/ct100684s} level of theory using the DFTB+ program~\cite{Aradi2007/10.1021/jp070186p}. The 3ob parameter set was used for the H and O atoms~\cite{Gaus2013/10.1021/ct300849w} which was combined with a recently introduced parameterization for the hydrated halide ion~\cite{Jahangir2013/10.1021/ct300919h}. Dispersion forces were included via a Lennard-Jones potential~\cite{Zhechkov2005/10.1021/ct050065y} with parameters taken from the Universal Force Field~\cite{Rappe1992/10.1021/ja00051a040}.

To identify the subensemble of protons connected to overcoordinated oxygen atoms we employed the nearest-neighbor criterion, assigning each proton to the nearest oxygen atom. Overcoordinated oxygen atoms were then identified as those that were triply coordinated.

\begin{acknowledgments}
We greatly thank William Carpenter and Andrei Tokmakoff for providing the experimental linear acid spectra and for helpful comments on the manuscript. This material is based upon work supported by the National Science Foundation under Grant No. CHE-1652960. T.E.M also acknowledges support from a Cottrell Scholarship from the Research Corporation for Science Advancement and the Camille Dreyfus Teacher-Scholar Awards Program. This research used resources of the National Energy Research Scientific Computing Center, a DOE Office of Science User Facility supported by the Office of Science of the U.S. Department of Energy under Contract No. DE-AC02-05CH11231. We would also like to thank Stanford University and the Stanford Research Computing Center for providing computational resources and support that have contributed to these research results. This work used the XStream computational resource, supported by the National Science Foundation Major Research Instrumentation program (ACI-1429830).
\end{acknowledgments}

\clearpage
\onecolumngrid

\appendix

\graphicspath{{figures_SI/}}

\setcounter{equation}{0}
\setcounter{figure}{0}
\makeatletter 
\renewcommand{\thefigure}{S\@arabic\c@figure}
\makeatother
\renewcommand{\theequation}{S\arabic{equation}}

\section*{Supporting Information}

\subsection{Vibrational density of states and its time evolution \label{sec:VDOS_time}}

The underlying dynamical timescales of the nuclei can be characterized by the vibrational density of states (VDOS). As illustrated in the main text, the VDOS encodes the features observed in the infra-red spectra, albeit with different relative intensities (Fig.~1).

In order to link vibrational motion to structural features, we calculated the time-dependent VDOS of individual protons in the following way: The velocity time series for a particular atom was multiplied by a symmetric Hann window function which was centered at a given point in time and that decays smoothly to zero. This segment of the velocity was then used to calculate the velocity autocorrelation function which was Fourier transformed to give the VDOS. This procedure yields a VDOS for the given atom that is semi-local around the chosen point in time. The process was then repeated for the whole trajectory for each hydrogen atom in the system, resulting in a time-dependent VDOS for the given atom. An example of such a VDOS is shown in Fig.~2 for a 100~ps segment of the trajectory of a hydrogen atom that was specifically chosen to illustrate the evolution of the spectrum into regions characteristic of proton defects. 

\subsection{Identification of the motions leading to the 1250~cm$^{-1}$ feature}

In order to gain insight into the type of vibrational motion that gives rise to the feature at 1250~cm$^{-1}$, we consider the spectrum of protons in defects where one of the hydrogen bonds is to a chloride ion. We use this as an example since, as shown in Fig.~3, the protons in the defect that point away from the chloride ion strongly exhibit the 1250~cm$^{-1}$ feature. This feature has previously been suggested to correspond to a proton shuttling motion. To assess this, we further decompose the spectrum of the protons that are not hydrogen bonded to the ion. We project their velocities onto the unit displacement vector between the oxygen atom to which they are covalently bound (i.e. the defect oxygen) and the second closest oxygen atom. 

\begin{figure}[ht]
  \centering
  \includegraphics[width=0.7\textwidth]{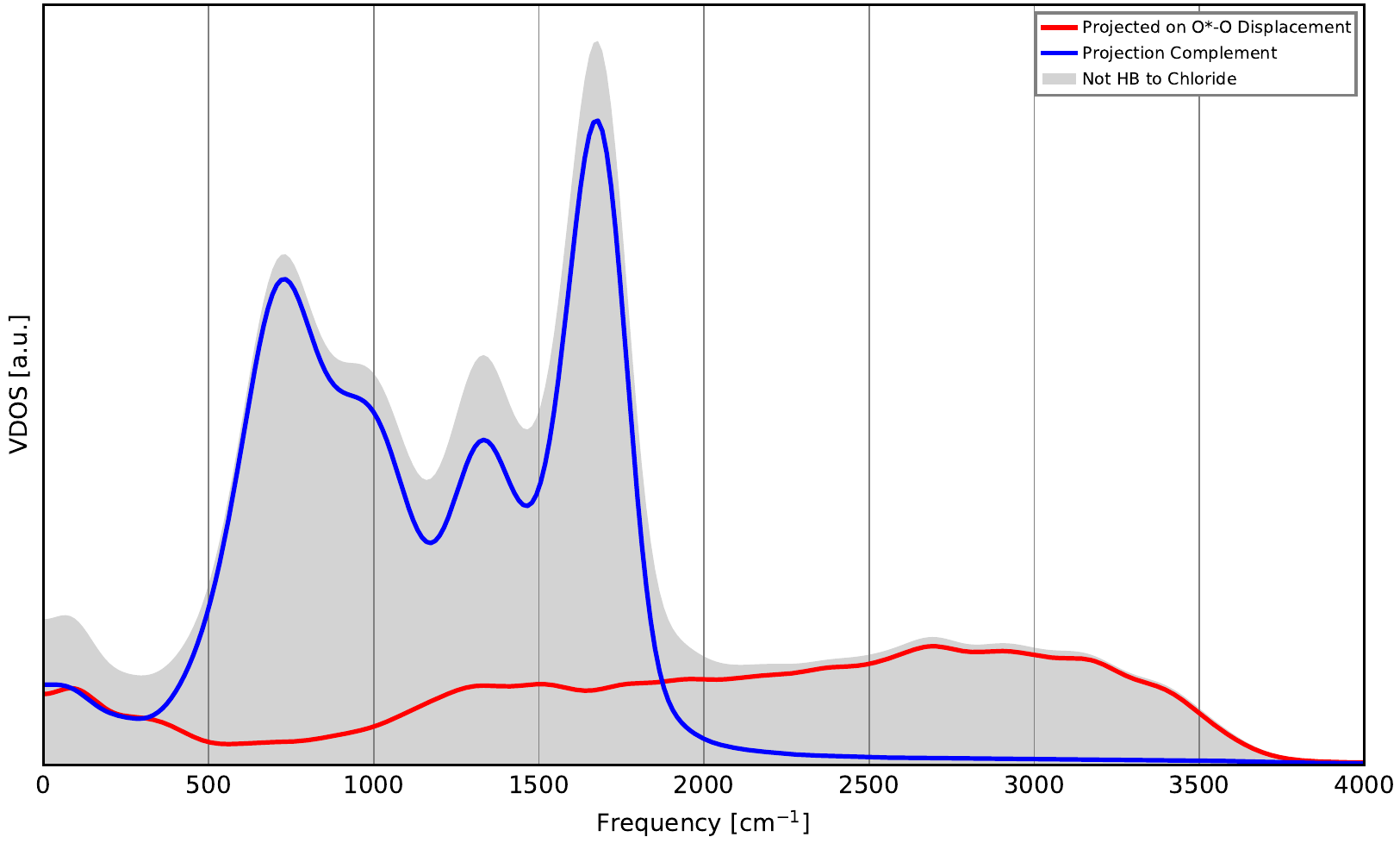}
  \caption{\label{fig:stretch_bend_decomp}
    Decomposition of the spectrum of the strongly shared protons. The VDOS obtained from velocities projected onto the O-O axis is shown in red, the VDOS obtained from the velocity perpendicular to this axis is shown blue. The VDOS of all protons in defects hydrogen-bonded to chloride ions that are not directly hydrogen bonded to the ion are shown as gray shading for reference.
  }
\end{figure}

As can be seen in Fig.~\ref{fig:stretch_bend_decomp}, the VDOS in direction of the oxygen-oxygen vector (red line) lacks the proton spectral features and instead contains just a broad intensity between the water bend and stretch frequencies. In contrast, the VDOS perpendicular to the oxygen-oxygen vector (blue line) exhibits the 1250~cm$^{-1}$ feature strongly, indicating that it arises from motion orthogonal to the O-O axis (i.e. a bending motion of the proton defect).

\subsection{Frequency time correlation functions}

\begin{figure}[ht]
  \centering
  \includegraphics[width=0.7\textwidth]{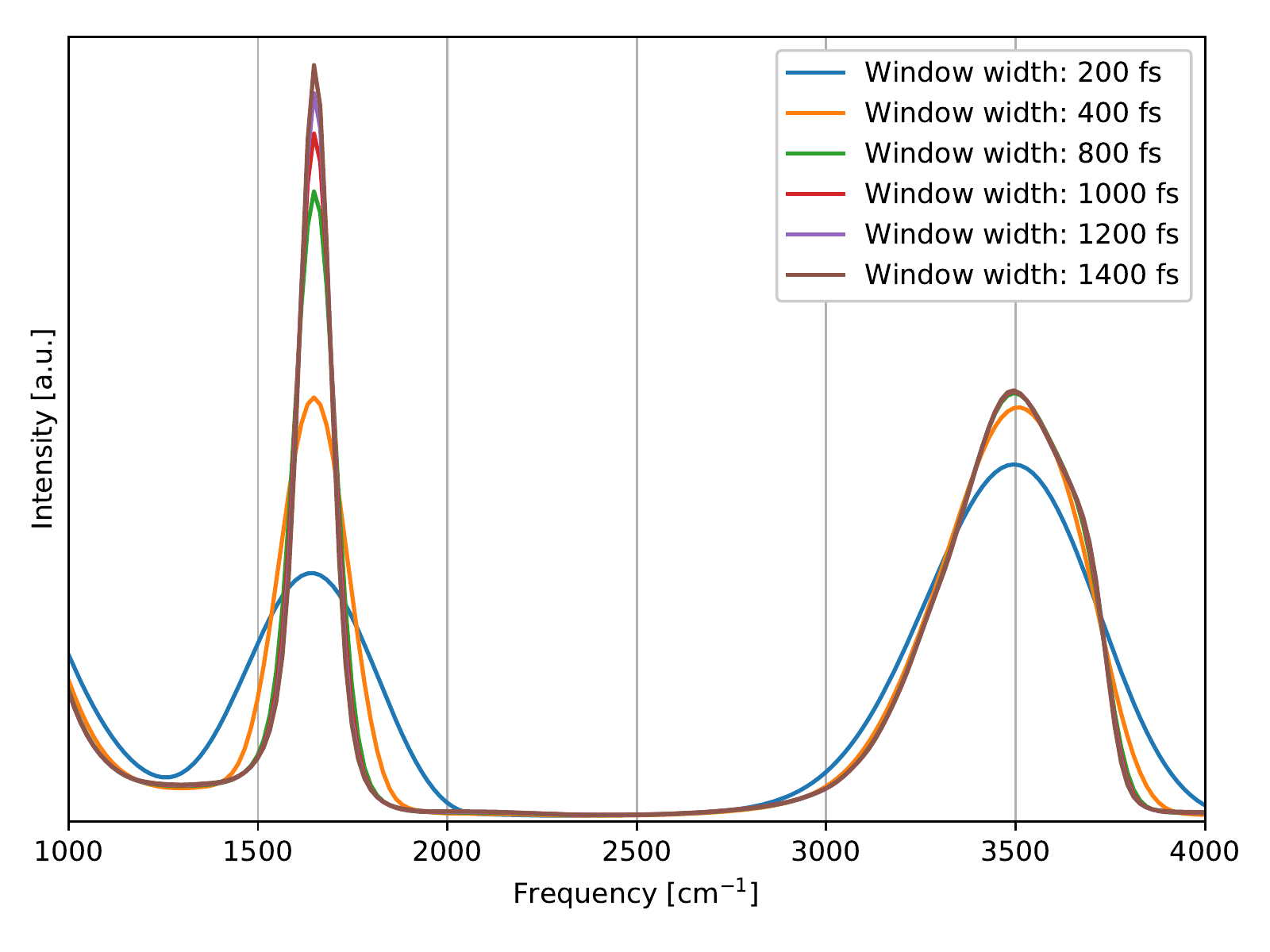}
  \caption{\label{fig:converge-VDOS}
    VDOS for all protons obtained from a classical ab initio molecular dynamics simulation of an aqueous excess proton defect as a function of the window width (dw). As the window width (given in the legend in femtoseconds) is increased the instantaneous spectrum at each frequency converges. 
  }
\end{figure}

The size of the data window can be chosen to balance the resolution in time and frequency of the time evolved spectrum. Fig.~\ref{fig:converge-VDOS} shows the convergence of the linear spectrum for the protons in a classical revPBE-D3 simulation of an aqueous excess proton defect. For a window width of $\sim$400~fs the stretch region at $\sim$3500~cm$^{-1}$ is converged while the bend region around 1600~cm$^{-1}$ is still artificially broadened by the windowing. With a window width of $\sim$800~fs both regions are converged.

\begin{figure}[ht]
  \centering
  \includegraphics[width=0.65\textwidth]{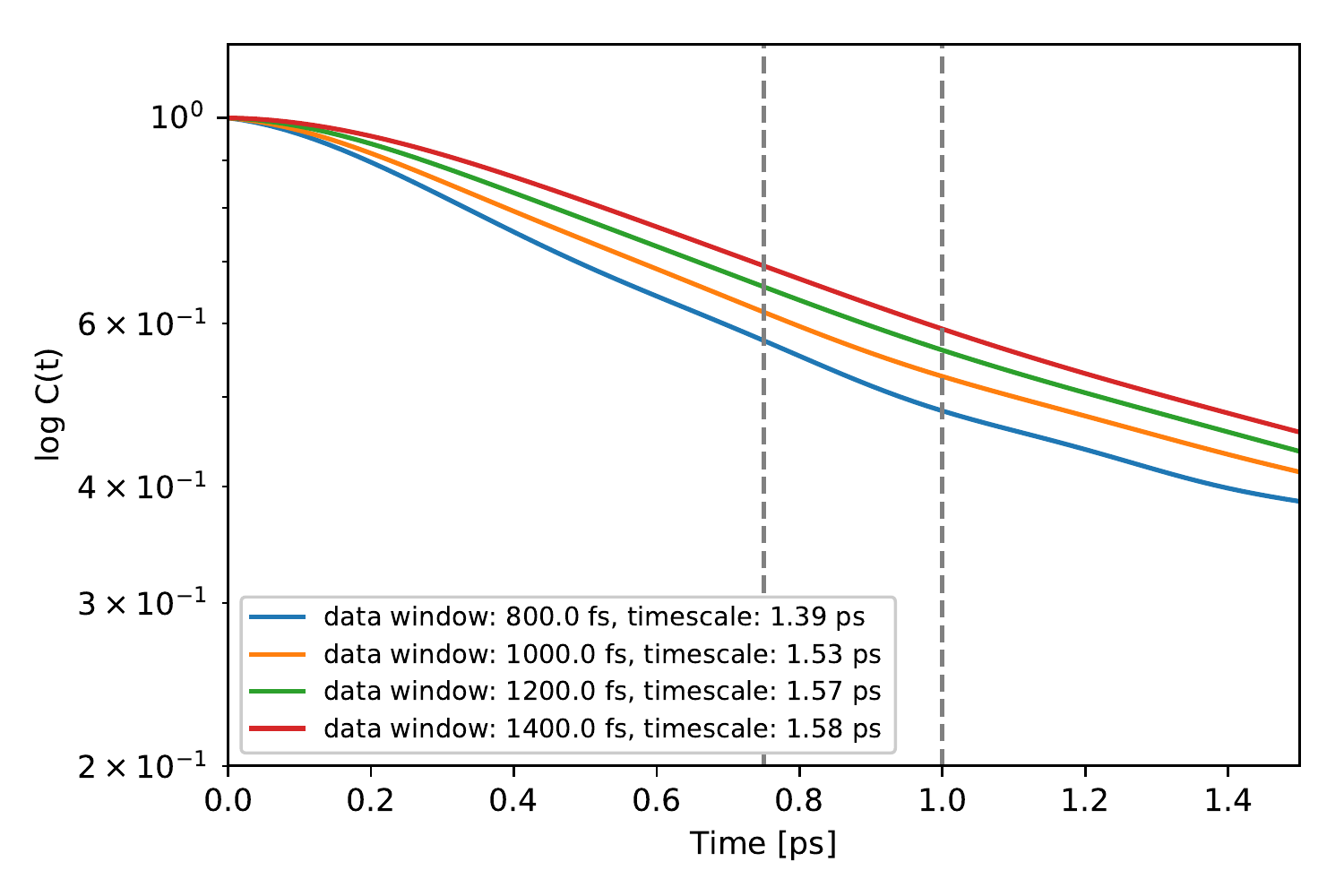}
  \caption{\label{fig:freq-cross-correlation}
    Frequency cross-correlation function obtained from our classical ab initio molecular dynamics simulation of an aqueous excess proton defect between 3150~cm$^{-1}$ and 1760~cm$^{-1}$. Note that the y-axis is on a log-scale. The colored lines show the correlation functions obtained using different data window sizes. 
    The time constants were obtained from a linear fit over the interval of the correlation function that is indicated by the gray vertical dashed lines (0.75--1.0~ps).
  }
\end{figure}

\begin{figure}[ht]
  \centering
  \includegraphics[width=0.65\textwidth]{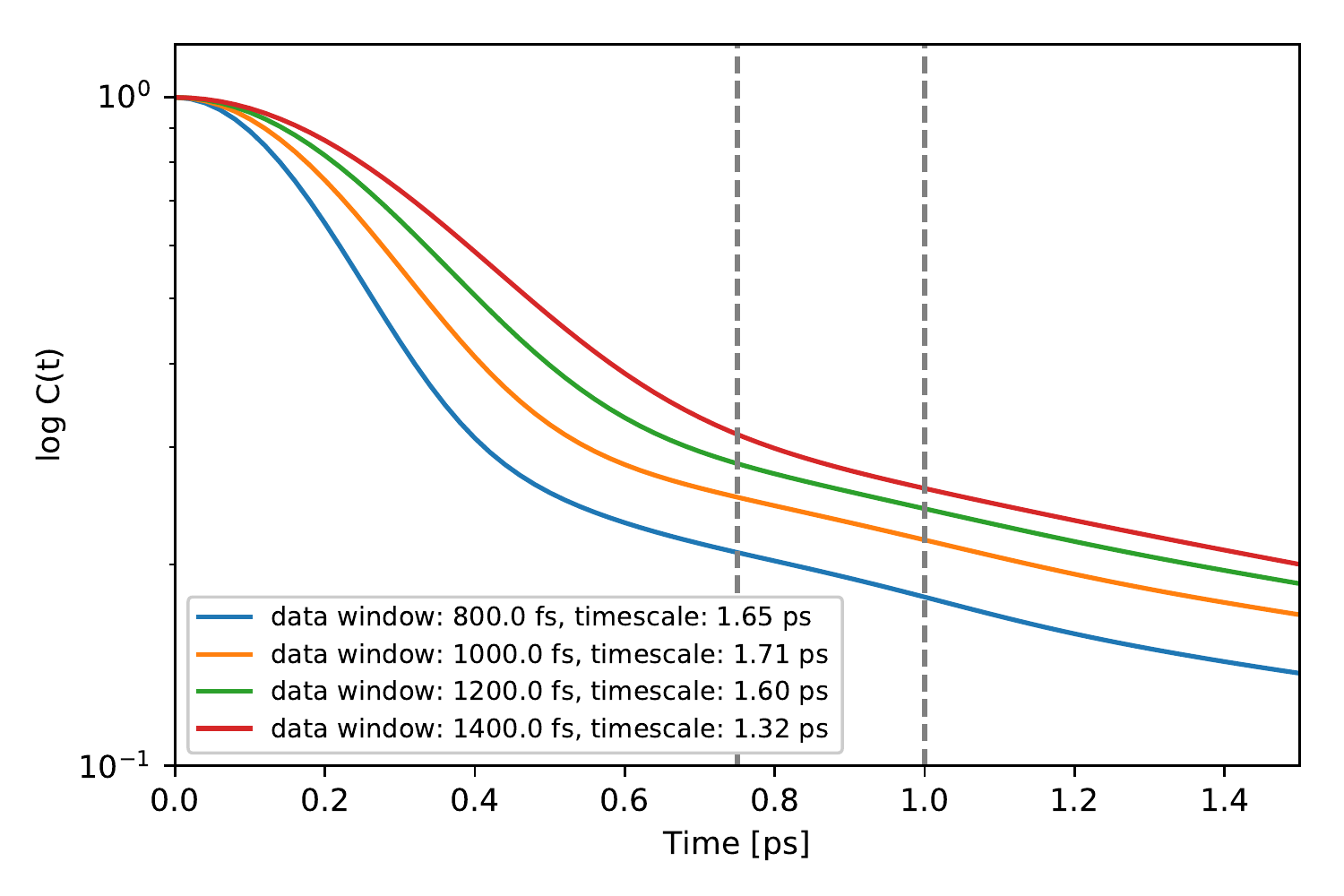}
  \caption{\label{fig:freq-self-correlation}
    Frequency autocorrelation function obtained from our classical ab initio molecular dynamics simulation of an aqueous excess proton defect for 1760~cm$^{-1}$. Note that the y-axis is on a log-scale. The colored lines show the correlation functions obtained using different data window sizes. The time constants were obtained from a linear fit over the interval of the correlation function that is indicated by the gray vertical dashed lines (0.75--1.0~ps).
  }
\end{figure}

Fig.~\ref{fig:freq-cross-correlation} shows the frequency-frequency time cross correlation function using the pump (3150~cm$^{-1}$) and probe (1760~cm$^{-1}$) frequencies studied in recent experiments while Fig.~\ref{fig:freq-self-correlation} shows the autocorrelation of the 1760~cm$^{-1}$ feature. The intensities of the frequencies at each point in time were computed using the windowing procedure for the VDOS discussed in Sec.~\ref{sec:VDOS_time}. As demonstrated in Fig.~\ref{fig:converge-VDOS} one can balance the resolution in time and frequency by changing the window width. The short time behavior of the correlation function is changed by the size of the window used while the long time decay, which occurs on timescales longer than the window width, is unaffected (see Fig.~\ref{fig:freq-cross-correlation} and Fig.~\ref{fig:freq-self-correlation}). More specifically, one expects there to be artifacts in the time correlation function arising from the windowing on a timescale of half the window width. Since we are interested in the slower processes occurring on the 0.5 to 1~ps timescale and the spectral region of interest (1760~cm$^{-1}$ -- 3150~cm$^{-1}$) converges with a window size of $\sim$800~fs the timescales extracted are robust with respect to this choice as shown in Fig.~\ref{fig:freq-cross-correlation} and Fig.~\ref{fig:freq-self-correlation}.

\subsection{Resolving the VDOS using the hydrogen bond balance}

\begin{figure}[ht]
  \centering
  \includegraphics[width=0.7\textwidth]{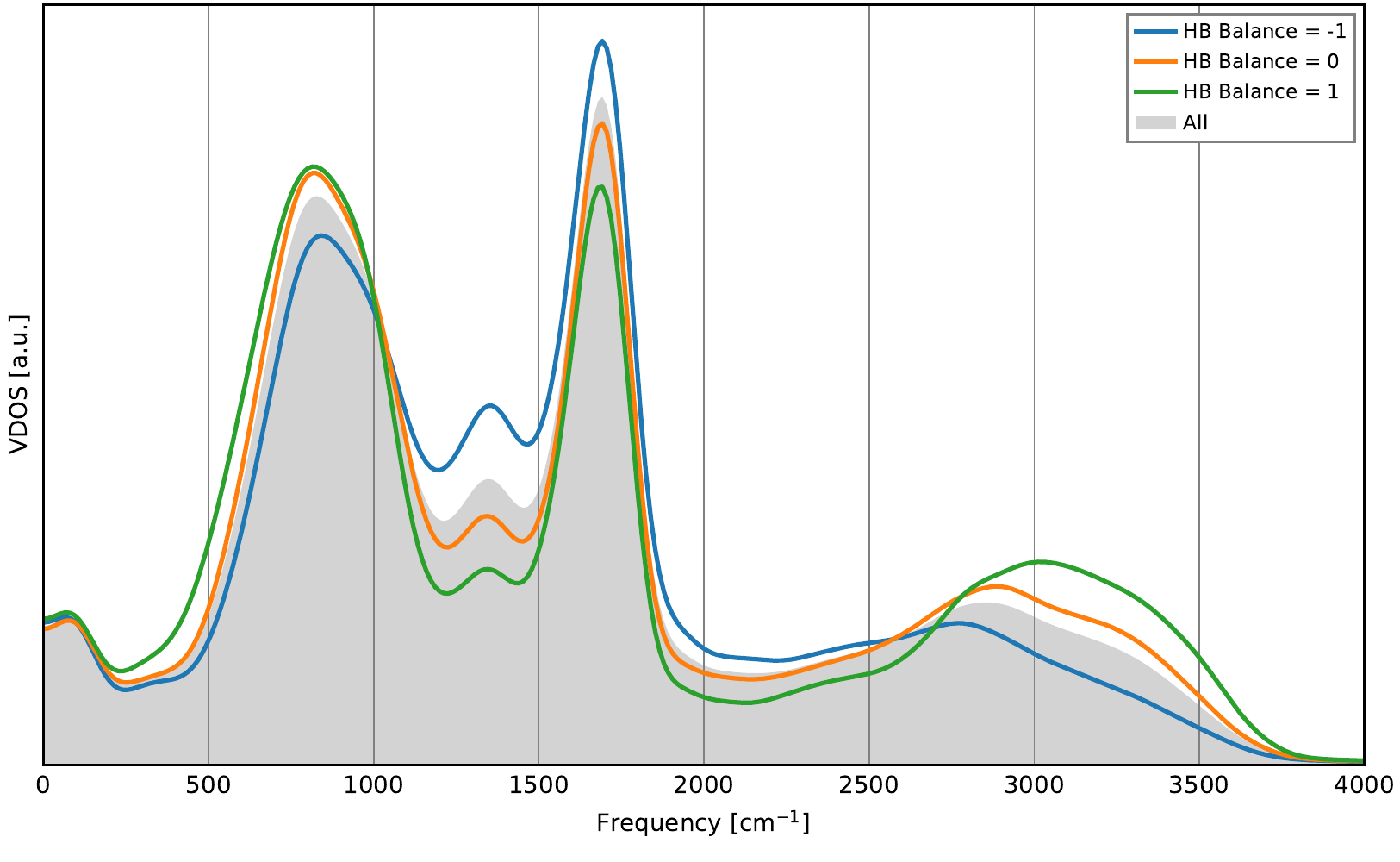}
  \caption{\label{fig:VDOS_balance_only}
    Proton defect VDOS resolved by the hydrogen bond balance. The VDOS of the defect protons is resolved by the three most frequently observed values of the hydrogen bond balance (-1,0,+1). Each spectrum is normalized to unity and the average spectrum across all three values of the hydrogen bond balance is shown as gray shading for reference. Data is shown for a classical revPBE-D3 simulation of an aqueous excess proton defect.
  }
\end{figure}

As shown in Fig.~5 in the main text, using the proton asymmetry coordinate $\phi$ allows the spectral features to be effectively separated. One could also consider doing this separation using just the hydrogen bond balance of the the acceptor water molecule which the proton is hydrogen bonded to. However, Fig.~\ref{fig:VDOS_balance_only} shows that using the hydrogen bond balance of the acceptor water molecule alone results in a much poorer separation of the proton spectral features. In particular, the feature around 1250~cm$^{-1}$ appears in the spectra corresponding to all values of the hydrogen bond balance. The hydrogen bond balance alone is thus a poor choice for analyzing the origins of the spectral features.

\subsection{Illustration of the $\phi$ coordinate}

\begin{figure}[ht]
  \centering
  \includegraphics[scale=0.7]{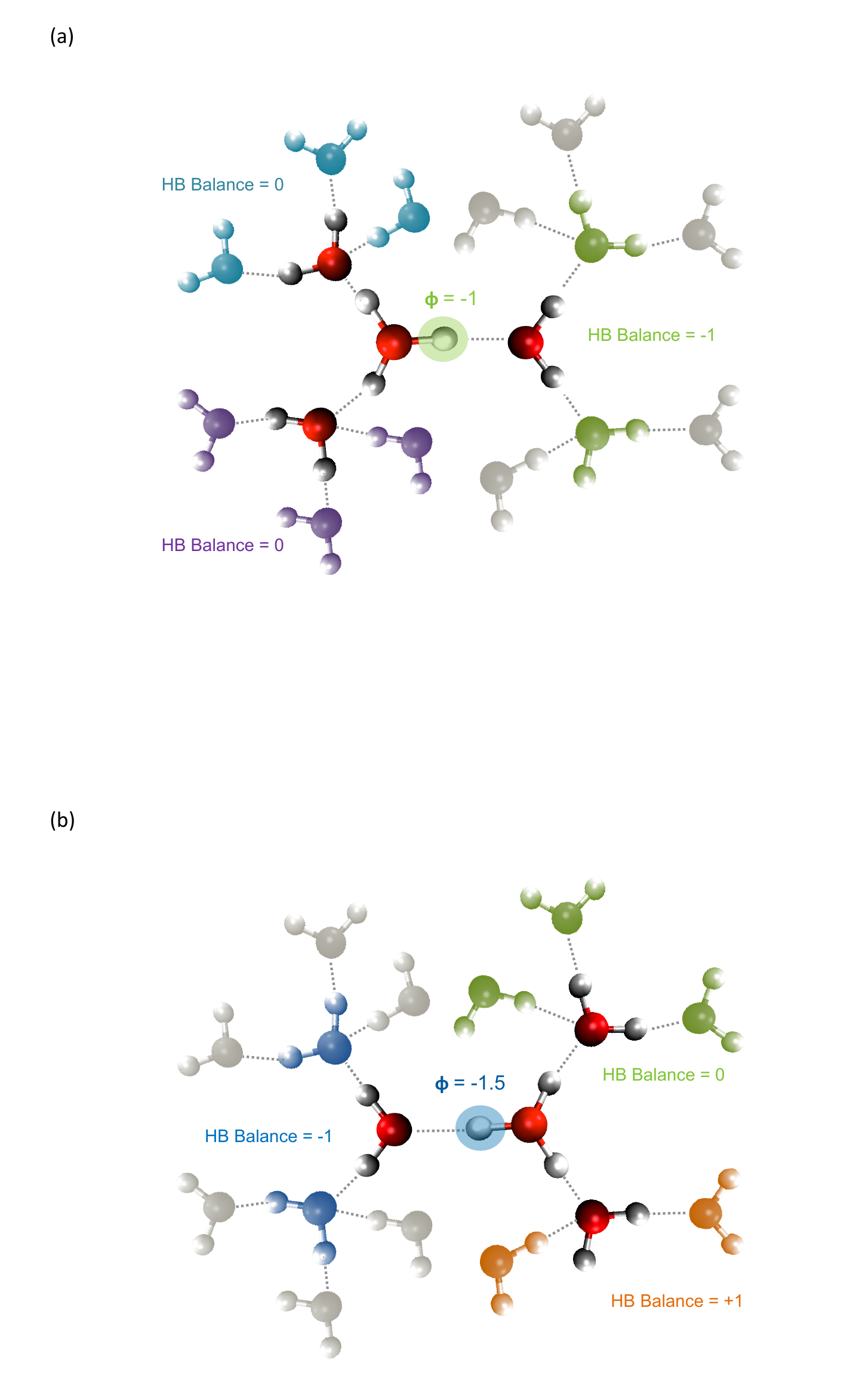}
  \caption{\label{fig:phi_illustration_zundel}
    Further illustration of the asymmetry coordinate $\phi$. The illustrations of $\phi$ are centered on a single proton (shaded in green and blue in the top and bottom panels, respectively) in order to emphasize that the calculation of $\phi$ remains well defined even in the event of proton transfer between two overcoordinated oxygens. The defect proton is shaded in the color that matches the environment it points at. In this example, proton transfer causes the $\phi$ value of the transferred proton to change from -1 to -1.5, as calculated from the colored environments. The gray waters are those which change from being active to inactive when calculating $\phi$ for the proton depending on the overcoordinated oxygen it is bound to.
  }
\end{figure}

Although the illustration in Fig. 4 showing how $\phi$ is defined for a proton could be seen as resembling a solvated ``Eigen'' structure, it is important to emphasize that the calculation of the asymmetry parameter $\phi$ does not assume anything about the structure of the proton defect in solution beyond the identification of overcoordinated oxygen atoms. In Fig.~\ref{fig:phi_illustration_zundel} we show that the definition of $\phi$ is also naturally compatible with a more ``Zundel''-centric picture where the proton is shared between two water molecules. Consider a case where one of the defect protons transfers between two oxygen atoms. The top and bottom panel of Fig.~\ref{fig:phi_illustration_zundel} show the situation where the proton is assigned to the left and right oxygen atom, respectively. In each case, the colored water molecules are used for the calculation of hydrogen bond balances and thus $\phi$ for the central proton, while the gray water molecules are shown for reference only.
From this it is clear that when the assignment of the proton changes to the other oxygen atom, $\phi$ is still defined for it (though it can change its $\phi$ value, as in the case illustrated in the figure), since it is still part of the defect. Hence, by having a $\phi$ parameter assigned to each proton, it remains well defined even for protons that are heavily shared or transfer frequently.
Thus, $\phi$ enables the quantification of the overall asymmetry of a proton's hydrogen bonding environment without biasing the analysis by assuming any particular structure of the proton defect.

\subsection{Correlation function and decorrelation time of the $\phi$ coordinate}

\begin{figure}[ht]
  \centering
  \includegraphics[width=0.7\textwidth]{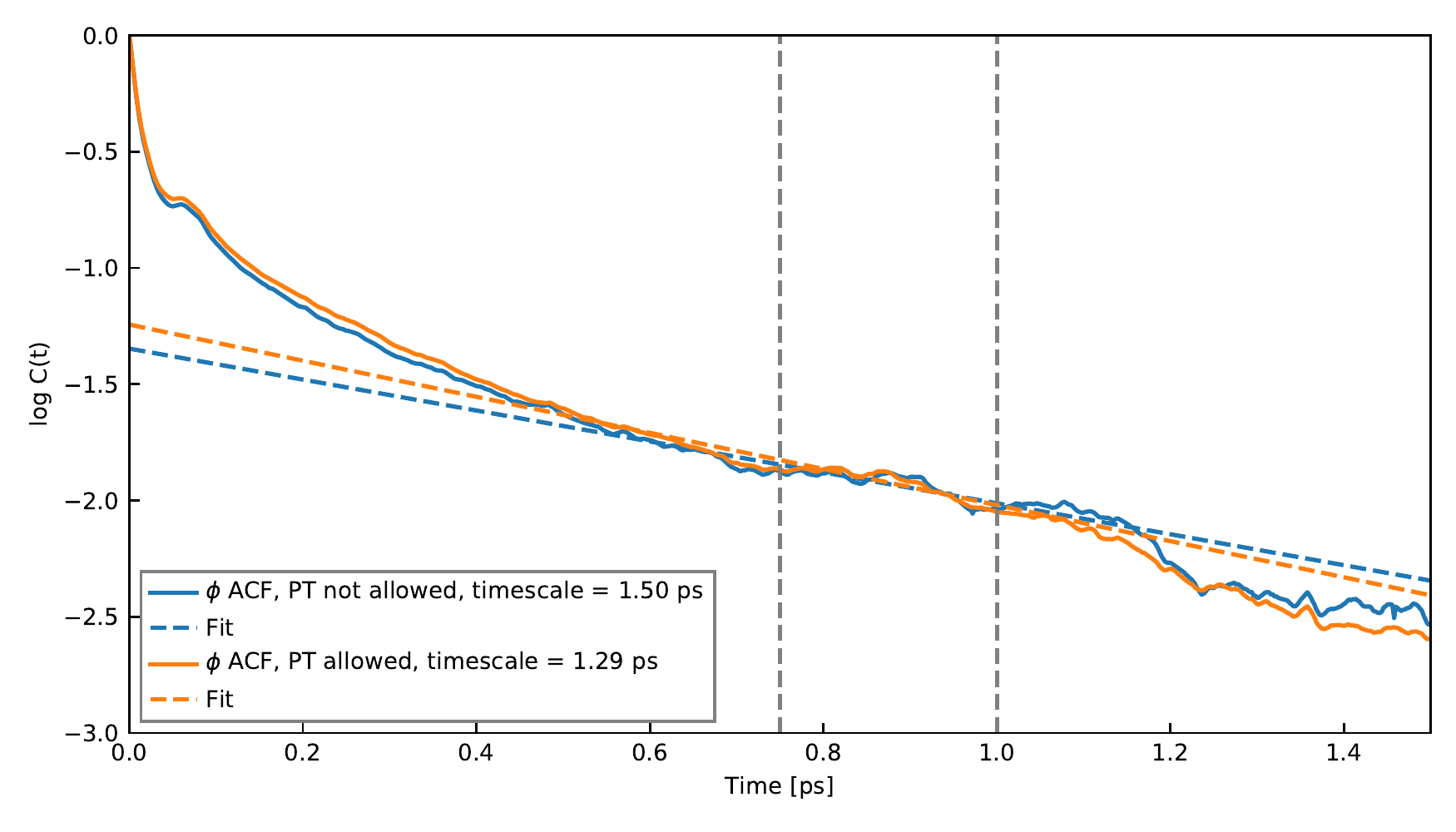}
  \caption{\label{fig:phi-acf}
    $\phi$ coordinate autocorrelation function obtained from our classical ab initio molecular dynamics  simulation of an aqueous excess proton defect. Note that the y-axis is on a log-scale. The orange line shows the correlation function when proton transfers are included and the blue line when they are not. The time constants were obtained from a linear fit (shown in blue and orange) over the interval of the correlation function that is indicated by the gray vertical dashed lines (0.75--1.0~ps).
  }
\end{figure}

Figure~\ref{fig:phi-acf} shows the logarithm of the autocorrelation function of the hydrogen bonding asymmetry coordinate, $\phi$, for the defect protons. The time constant extracted from a fit to the interval indicated by vertical dashed gray lines is $\sim$1.29~ps. The orange line shows the $\phi$ correlation function allowing for contributions whenever a proton is part of a proton defect (i.e. if a proton transfers to a new overcoordinated oxygen it still contributes to the correlation function). In contrast, the blue line shows the $\phi$ correlation where the proton does not contribute to the correlation function when it is not connected to its original (t=0) overcoordinated oxygen. The similarity of these two correlation functions indicates that proton transfer, which involves a change in the identity of the overcoordinated oxygen, plays a small role in decorrelating the $\phi$ coordinate.

\subsection{Correlation function and decorrelation time of the proton sharing coordinate $\delta$}

\begin{figure}[ht]
  \centering
  \includegraphics[width=0.7\textwidth]{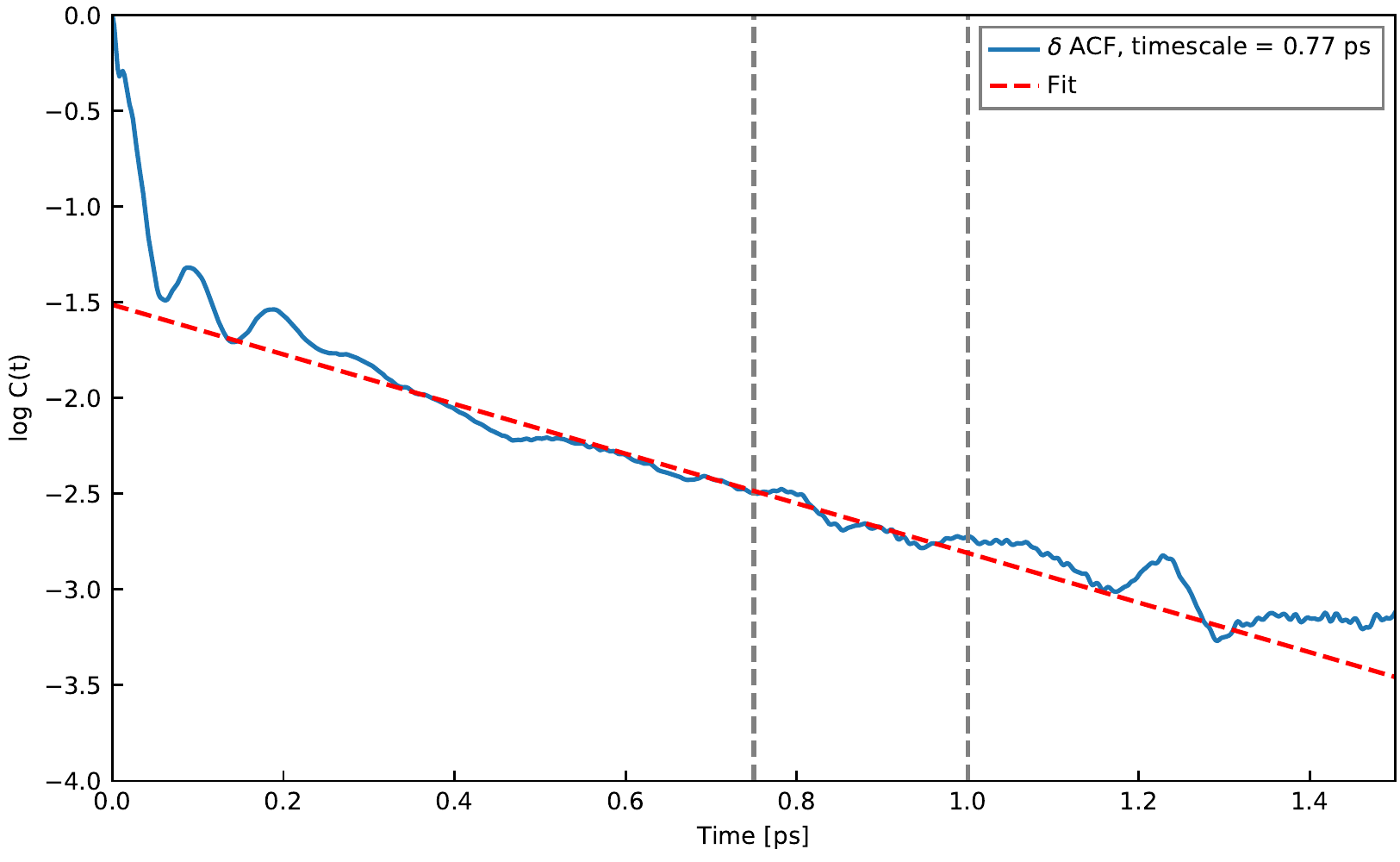}
  \caption{\label{fig:delta-acf}
     $\delta$-coordinate autocorrelation function for defect protons obtained from our classical ab initio molecular dynamics simulation of an aqueous excess proton defect. Note that the y-axis is on a log-scale. An initial rapid decorrelation of the coordinate in $\sim$100~fs is followed by a slower decay.  The time constant was obtained from a linear fit (shown in red) over the interval of the correlation function that is indicated by the gray vertical dashed lines (0.75--1.0~ps).
  }
\end{figure}

Figure~\ref{fig:delta-acf} shows the logarithm of the autocorrelation function of the proton sharing coordinate $\delta$ for the defect protons in a classical revPBE-D3 simulation of an aqueous excess proton defect. All protons bound to the overcoordinated oxygen at a given point in time contribute to the correlation function. The autocorrelation function exhibits a rapid initial decay to $\sim$0.2 in under 100~fs, followed by a slower decay to 0. The slower time scale obtained from a linear fit to the logarithm of the correlation function is 0.77~ps.

\subsection{Convergence of Ring Polymer Contraction}

\begin{figure}[ht]
  \centering
  \includegraphics[width=0.6\textwidth]{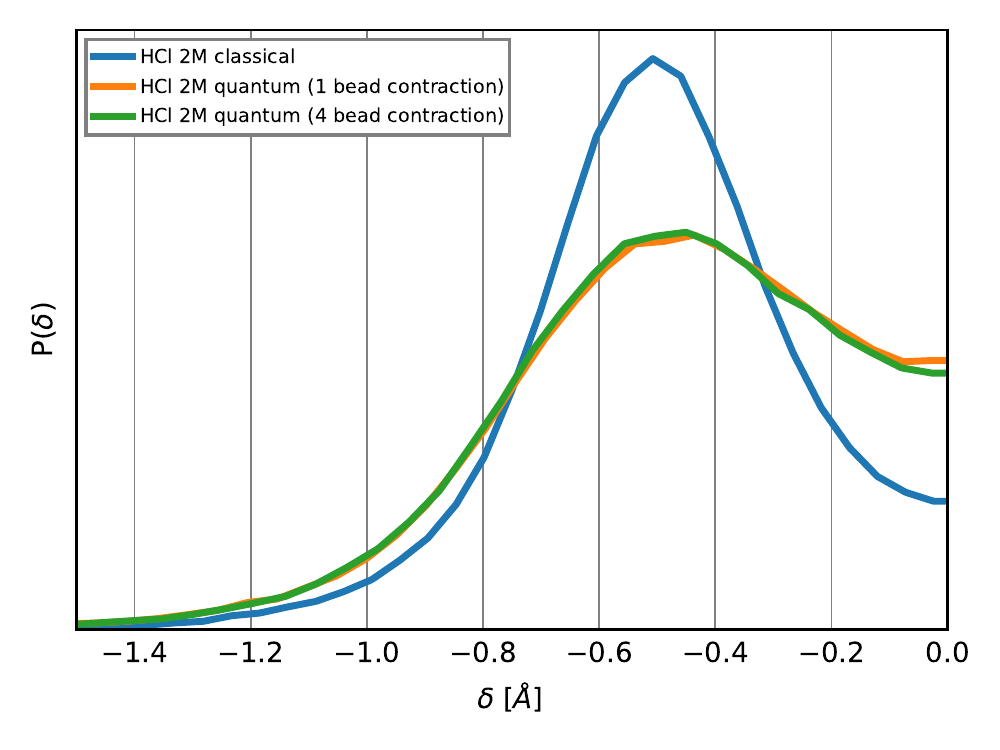}
  \caption{\label{fig:contraction-convergence}
    Ring polymer contraction convergence. The probability distributions of the proton sharing coordinate for H atoms connected to O* is shown for ring polymer contraction to $P'=1$ (centroid contraction) and $P'=4$ replicas.
  }
\end{figure}

Path integral calculations were performed using ring polymer contraction (RPC) to $P'=1$ replicas using an MTS propagator with an outer time step of 2.0~fs and an inner time step of 0.25~fs. Figure~\ref{fig:contraction-convergence} illustrates that simulations employing RPC to $P'=4$ replicas yield results graphically indistinguishable to the result for RPC to $P'=1$ for the probability distribution along the proton sharing coordinate which is one of the properties most sensitive to nuclear quantum effects. This indicates that $P'=1$ is converged using the SCC-DFTB reference. 

\begin{thebibliography}{74}%
\makeatletter
\providecommand \@ifxundefined [1]{%
 \@ifx{#1\undefined}
}%
\providecommand \@ifnum [1]{%
 \ifnum #1\expandafter \@firstoftwo
 \else \expandafter \@secondoftwo
 \fi
}%
\providecommand \@ifx [1]{%
 \ifx #1\expandafter \@firstoftwo
 \else \expandafter \@secondoftwo
 \fi
}%
\providecommand \natexlab [1]{#1}%
\providecommand \enquote  [1]{``#1''}%
\providecommand \bibnamefont  [1]{#1}%
\providecommand \bibfnamefont [1]{#1}%
\providecommand \citenamefont [1]{#1}%
\providecommand \href@noop [0]{\@secondoftwo}%
\providecommand \href [0]{\begingroup \@sanitize@url \@href}%
\providecommand \@href[1]{\@@startlink{#1}\@@href}%
\providecommand \@@href[1]{\endgroup#1\@@endlink}%
\providecommand \@sanitize@url [0]{\catcode `\\12\catcode `\$12\catcode
  `\&12\catcode `\#12\catcode `\^12\catcode `\_12\catcode `\%12\relax}%
\providecommand \@@startlink[1]{}%
\providecommand \@@endlink[0]{}%
\providecommand \url  [0]{\begingroup\@sanitize@url \@url }%
\providecommand \@url [1]{\endgroup\@href {#1}{\urlprefix }}%
\providecommand \urlprefix  [0]{URL }%
\providecommand \Eprint [0]{\href }%
\providecommand \doibase [0]{http://dx.doi.org/}%
\providecommand \selectlanguage [0]{\@gobble}%
\providecommand \bibinfo  [0]{\@secondoftwo}%
\providecommand \bibfield  [0]{\@secondoftwo}%
\providecommand \translation [1]{[#1]}%
\providecommand \BibitemOpen [0]{}%
\providecommand \bibitemStop [0]{}%
\providecommand \bibitemNoStop [0]{.\EOS\space}%
\providecommand \EOS [0]{\spacefactor3000\relax}%
\providecommand \BibitemShut  [1]{\csname bibitem#1\endcsname}%
\let\auto@bib@innerbib\@empty
\bibitem [{\citenamefont {Kreuer}(1996)}]{Kreuer1996/10.1021/cm950192a}%
  \BibitemOpen
  \bibfield  {author} {\bibinfo {author} {\bibfnamefont {K.-D.}\ \bibnamefont
  {Kreuer}},\ }\href {\doibase 10.1021/cm950192a} {\bibfield  {journal}
  {\bibinfo  {journal} {Chemistry of Materials}\ }\textbf {\bibinfo {volume}
  {8}},\ \bibinfo {pages} {610} (\bibinfo {year} {1996})}\BibitemShut {NoStop}%
\bibitem [{\citenamefont
  {Decoursey}(2003)}]{Decoursey2003/10.1152/physrev.00028.2002}%
  \BibitemOpen
  \bibfield  {author} {\bibinfo {author} {\bibfnamefont {T.}~\bibnamefont
  {Decoursey}},\ }\href {\doibase 10.1152/physrev.00028.2002} {\bibfield
  {journal} {\bibinfo  {journal} {Physiological reviews}\ }\textbf {\bibinfo
  {volume} {83}},\ \bibinfo {pages} {475} (\bibinfo {year} {2003})}\BibitemShut
  {NoStop}%
\bibitem [{\citenamefont
  {Wraight}(2006)}]{Wraight2006/10.1016/j.bbabio.2006.06.017}%
  \BibitemOpen
  \bibfield  {author} {\bibinfo {author} {\bibfnamefont {C.~A.}\ \bibnamefont
  {Wraight}},\ }\href {\doibase 10.1016/j.bbabio.2006.06.017} {\bibfield
  {journal} {\bibinfo  {journal} {Biochimica et Biophysica Acta -
  Bioenergetics}\ }\textbf {\bibinfo {volume} {1757}},\ \bibinfo {pages} {886}
  (\bibinfo {year} {2006})}\BibitemShut {NoStop}%
\bibitem [{\citenamefont {Peighambardoust}, \citenamefont {Rowshanzamir},\ and\
  \citenamefont
  {Amjadi}(2010)}]{Peighambardoust2010/10.1016/j.ijhydene.2010.05.017}%
  \BibitemOpen
  \bibfield  {author} {\bibinfo {author} {\bibfnamefont {S.~J.}\ \bibnamefont
  {Peighambardoust}}, \bibinfo {author} {\bibfnamefont {S.}~\bibnamefont
  {Rowshanzamir}}, \ and\ \bibinfo {author} {\bibfnamefont {M.}~\bibnamefont
  {Amjadi}},\ }\href {\doibase 10.1016/j.ijhydene.2010.05.017} {\emph {\bibinfo
  {title} {International Journal of Hydrogen Energy}}},\ Vol.~\bibinfo {volume}
  {35}\ (\bibinfo  {publisher} {Elsevier Ltd},\ \bibinfo {year} {2010})\ pp.\
  \bibinfo {pages} {9349--9384}\BibitemShut {NoStop}%
\bibitem [{\citenamefont {Headrick}, \citenamefont {Bopp},\ and\ \citenamefont
  {Johnson}(2004)}]{Headrick2004/10.1063/1.1834566}%
  \BibitemOpen
  \bibfield  {author} {\bibinfo {author} {\bibfnamefont {J.~M.}\ \bibnamefont
  {Headrick}}, \bibinfo {author} {\bibfnamefont {J.~C.}\ \bibnamefont {Bopp}},
  \ and\ \bibinfo {author} {\bibfnamefont {M.~A.}\ \bibnamefont {Johnson}},\
  }\href {\doibase 10.1063/1.1834566} {\bibfield  {journal} {\bibinfo
  {journal} {Journal of Chemical Physics}\ }\textbf {\bibinfo {volume} {121}},\
  \bibinfo {pages} {11523} (\bibinfo {year} {2004})}\BibitemShut {NoStop}%
\bibitem [{\citenamefont {Headrick}\ \emph {et~al.}(2005)\citenamefont
  {Headrick}, \citenamefont {Diken}, \citenamefont {Walters}, \citenamefont
  {Hammer}, \citenamefont {Christie}, \citenamefont {Cui}, \citenamefont
  {Myshakin}, \citenamefont {Duncan}, \citenamefont {Johnson},\ and\
  \citenamefont {Jordan}}]{Headrick2005/10.1126/science.1113094}%
  \BibitemOpen
  \bibfield  {author} {\bibinfo {author} {\bibfnamefont {J.~M.}\ \bibnamefont
  {Headrick}}, \bibinfo {author} {\bibfnamefont {E.~G.}\ \bibnamefont {Diken}},
  \bibinfo {author} {\bibfnamefont {R.~S.}\ \bibnamefont {Walters}}, \bibinfo
  {author} {\bibfnamefont {N.~I.}\ \bibnamefont {Hammer}}, \bibinfo {author}
  {\bibfnamefont {R.~A.}\ \bibnamefont {Christie}}, \bibinfo {author}
  {\bibfnamefont {J.}~\bibnamefont {Cui}}, \bibinfo {author} {\bibfnamefont
  {E.~M.}\ \bibnamefont {Myshakin}}, \bibinfo {author} {\bibfnamefont {M.~A.}\
  \bibnamefont {Duncan}}, \bibinfo {author} {\bibfnamefont {M.~A.}\
  \bibnamefont {Johnson}}, \ and\ \bibinfo {author} {\bibfnamefont {K.~D.}\
  \bibnamefont {Jordan}},\ }\href {\doibase 10.1126/science.1113094} {\bibfield
   {journal} {\bibinfo  {journal} {Science}\ }\textbf {\bibinfo {volume}
  {308}},\ \bibinfo {pages} {1765} (\bibinfo {year} {2005})}\BibitemShut
  {NoStop}%
\bibitem [{\citenamefont {Woutersen}\ and\ \citenamefont
  {Bakker}(2006)}]{Woutersen2006/10.1103/PhysRevLett.96.138305}%
  \BibitemOpen
  \bibfield  {author} {\bibinfo {author} {\bibfnamefont {S.}~\bibnamefont
  {Woutersen}}\ and\ \bibinfo {author} {\bibfnamefont {H.~J.}\ \bibnamefont
  {Bakker}},\ }\href {\doibase 10.1103/PhysRevLett.96.138305} {\bibfield
  {journal} {\bibinfo  {journal} {Physical Review Letters}\ }\textbf {\bibinfo
  {volume} {96}},\ \bibinfo {pages} {5} (\bibinfo {year} {2006})}\BibitemShut
  {NoStop}%
\bibitem [{\citenamefont {Thamer}\ \emph {et~al.}(2015)\citenamefont {Thamer},
  \citenamefont {{De Marco}}, \citenamefont {Ramasesha}, \citenamefont
  {Mandal},\ and\ \citenamefont
  {Tokmakoff}}]{Thamer2015/10.1126/science.aab3908}%
  \BibitemOpen
  \bibfield  {author} {\bibinfo {author} {\bibfnamefont {M.}~\bibnamefont
  {Thamer}}, \bibinfo {author} {\bibfnamefont {L.}~\bibnamefont {{De Marco}}},
  \bibinfo {author} {\bibfnamefont {K.}~\bibnamefont {Ramasesha}}, \bibinfo
  {author} {\bibfnamefont {A.}~\bibnamefont {Mandal}}, \ and\ \bibinfo {author}
  {\bibfnamefont {A.}~\bibnamefont {Tokmakoff}},\ }\href {\doibase
  10.1126/science.aab3908} {\bibfield  {journal} {\bibinfo  {journal}
  {Science}\ }\textbf {\bibinfo {volume} {350}},\ \bibinfo {pages} {78}
  (\bibinfo {year} {2015})}\BibitemShut {NoStop}%
\bibitem [{\citenamefont {Dahms}\ \emph {et~al.}(2016)\citenamefont {Dahms},
  \citenamefont {Costard}, \citenamefont {Pines}, \citenamefont {Fingerhut},
  \citenamefont {Nibbering},\ and\ \citenamefont
  {Elsaesser}}]{Dahms2016/10.1002/anie.201602523}%
  \BibitemOpen
  \bibfield  {author} {\bibinfo {author} {\bibfnamefont {F.}~\bibnamefont
  {Dahms}}, \bibinfo {author} {\bibfnamefont {R.}~\bibnamefont {Costard}},
  \bibinfo {author} {\bibfnamefont {E.}~\bibnamefont {Pines}}, \bibinfo
  {author} {\bibfnamefont {B.~P.}\ \bibnamefont {Fingerhut}}, \bibinfo {author}
  {\bibfnamefont {E.~T.~J.}\ \bibnamefont {Nibbering}}, \ and\ \bibinfo
  {author} {\bibfnamefont {T.}~\bibnamefont {Elsaesser}},\ }\href {\doibase
  10.1002/anie.201602523} {\bibfield  {journal} {\bibinfo  {journal}
  {Angewandte Chemie - International Edition}\ }\textbf {\bibinfo {volume}
  {55}},\ \bibinfo {pages} {10600} (\bibinfo {year} {2016})}\BibitemShut
  {NoStop}%
\bibitem [{\citenamefont {Agmon}\ \emph {et~al.}(2016)\citenamefont {Agmon},
  \citenamefont {Bakker}, \citenamefont {Campen}, \citenamefont {Henchman},
  \citenamefont {Pohl}, \citenamefont {Roke}, \citenamefont {Th??mer},\ and\
  \citenamefont {Hassanali}}]{Agmon2016/10.1021/acs.chemrev.5b00736}%
  \BibitemOpen
  \bibfield  {author} {\bibinfo {author} {\bibfnamefont {N.}~\bibnamefont
  {Agmon}}, \bibinfo {author} {\bibfnamefont {H.~J.}\ \bibnamefont {Bakker}},
  \bibinfo {author} {\bibfnamefont {R.~K.}\ \bibnamefont {Campen}}, \bibinfo
  {author} {\bibfnamefont {R.~H.}\ \bibnamefont {Henchman}}, \bibinfo {author}
  {\bibfnamefont {P.}~\bibnamefont {Pohl}}, \bibinfo {author} {\bibfnamefont
  {S.}~\bibnamefont {Roke}}, \bibinfo {author} {\bibfnamefont {M.}~\bibnamefont
  {Th??mer}}, \ and\ \bibinfo {author} {\bibfnamefont {A.}~\bibnamefont
  {Hassanali}},\ }\href {\doibase 10.1021/acs.chemrev.5b00736} {\bibfield
  {journal} {\bibinfo  {journal} {Chemical Reviews}\ }\textbf {\bibinfo
  {volume} {116}},\ \bibinfo {pages} {7642} (\bibinfo {year}
  {2016})}\BibitemShut {NoStop}%
\bibitem [{\citenamefont {Tuckerman}\ \emph {et~al.}(1995)\citenamefont
  {Tuckerman}, \citenamefont {Laasonen}, \citenamefont {Sprik},\ and\
  \citenamefont {Parrinello}}]{Tuckerman1995/10.1021/j100016a003}%
  \BibitemOpen
  \bibfield  {author} {\bibinfo {author} {\bibfnamefont {M.}~\bibnamefont
  {Tuckerman}}, \bibinfo {author} {\bibfnamefont {K.}~\bibnamefont {Laasonen}},
  \bibinfo {author} {\bibfnamefont {M.}~\bibnamefont {Sprik}}, \ and\ \bibinfo
  {author} {\bibfnamefont {M.}~\bibnamefont {Parrinello}},\ }\href {\doibase
  10.1021/j100016a003} {\bibfield  {journal} {\bibinfo  {journal} {The Journal
  of Physical Chemistry}\ }\textbf {\bibinfo {volume} {99}},\ \bibinfo {pages}
  {5749} (\bibinfo {year} {1995})}\BibitemShut {NoStop}%
\bibitem [{\citenamefont {Marx}\ \emph {et~al.}(1999)\citenamefont {Marx},
  \citenamefont {Tuckerman}, \citenamefont {Hutter},\ and\ \citenamefont
  {Parrinello}}]{Marx1999/10.1038/17579}%
  \BibitemOpen
  \bibfield  {author} {\bibinfo {author} {\bibfnamefont {D.}~\bibnamefont
  {Marx}}, \bibinfo {author} {\bibfnamefont {M.~E.}\ \bibnamefont {Tuckerman}},
  \bibinfo {author} {\bibfnamefont {J.}~\bibnamefont {Hutter}}, \ and\ \bibinfo
  {author} {\bibfnamefont {M.}~\bibnamefont {Parrinello}},\ }\href {\doibase
  10.1038/17579} {\bibfield  {journal} {\bibinfo  {journal} {Nature}\ }\textbf
  {\bibinfo {volume} {397}},\ \bibinfo {pages} {601} (\bibinfo {year}
  {1999})}\BibitemShut {NoStop}%
\bibitem [{\citenamefont {Iftimie}\ and\ \citenamefont
  {Tuckerman}(2006)}]{Iftimie2006/10.1002/anie.200502259}%
  \BibitemOpen
  \bibfield  {author} {\bibinfo {author} {\bibfnamefont {R.}~\bibnamefont
  {Iftimie}}\ and\ \bibinfo {author} {\bibfnamefont {M.~E.}\ \bibnamefont
  {Tuckerman}},\ }\href {\doibase 10.1002/anie.200502259} {\bibfield  {journal}
  {\bibinfo  {journal} {Angewandte Chemie - International Edition}\ }\textbf
  {\bibinfo {volume} {45}},\ \bibinfo {pages} {1144} (\bibinfo {year}
  {2006})}\BibitemShut {NoStop}%
\bibitem [{\citenamefont {Kaledin}, \citenamefont {Kaledin},\ and\
  \citenamefont {Bowman}(2006)}]{Kaledin2006/10.1021/jp054374w}%
  \BibitemOpen
  \bibfield  {author} {\bibinfo {author} {\bibfnamefont {M.}~\bibnamefont
  {Kaledin}}, \bibinfo {author} {\bibfnamefont {A.~L.}\ \bibnamefont
  {Kaledin}}, \ and\ \bibinfo {author} {\bibfnamefont {J.~M.}\ \bibnamefont
  {Bowman}},\ }\href {\doibase 10.1021/jp054374w} {\bibfield  {journal}
  {\bibinfo  {journal} {The Journal of Physical Chemistry A}\ }\textbf
  {\bibinfo {volume} {110}},\ \bibinfo {pages} {2933} (\bibinfo {year}
  {2006})}\BibitemShut {NoStop}%
\bibitem [{\citenamefont {Chandra}, \citenamefont {Tuckerman},\ and\
  \citenamefont {Marx}(2007)}]{Chandra2007/10.1103/PhysRevLett.99.145901}%
  \BibitemOpen
  \bibfield  {author} {\bibinfo {author} {\bibfnamefont {A.}~\bibnamefont
  {Chandra}}, \bibinfo {author} {\bibfnamefont {M.~E.}\ \bibnamefont
  {Tuckerman}}, \ and\ \bibinfo {author} {\bibfnamefont {D.}~\bibnamefont
  {Marx}},\ }\href {\doibase 10.1103/PhysRevLett.99.145901} {\bibfield
  {journal} {\bibinfo  {journal} {Physical Review Letters}\ }\textbf {\bibinfo
  {volume} {99}},\ \bibinfo {pages} {1} (\bibinfo {year} {2007})}\BibitemShut
  {NoStop}%
\bibitem [{\citenamefont {Xu}, \citenamefont {Izvekov},\ and\ \citenamefont
  {Voth}(2010)}]{Xu2010/10.1021/jp102516h}%
  \BibitemOpen
  \bibfield  {author} {\bibinfo {author} {\bibfnamefont {J.}~\bibnamefont
  {Xu}}, \bibinfo {author} {\bibfnamefont {S.}~\bibnamefont {Izvekov}}, \ and\
  \bibinfo {author} {\bibfnamefont {G.~A.}\ \bibnamefont {Voth}},\ }\href
  {\doibase 10.1021/jp102516h} {\bibfield  {journal} {\bibinfo  {journal}
  {Journal of Physical Chemistry B}\ }\textbf {\bibinfo {volume} {114}},\
  \bibinfo {pages} {9555} (\bibinfo {year} {2010})}\BibitemShut {NoStop}%
\bibitem [{\citenamefont {Xu}, \citenamefont {Zhang},\ and\ \citenamefont
  {Voth}(2011)}]{Xu2011/10.1021/jz101536b}%
  \BibitemOpen
  \bibfield  {author} {\bibinfo {author} {\bibfnamefont {J.}~\bibnamefont
  {Xu}}, \bibinfo {author} {\bibfnamefont {Y.}~\bibnamefont {Zhang}}, \ and\
  \bibinfo {author} {\bibfnamefont {G.~A.}\ \bibnamefont {Voth}},\ }\href
  {\doibase 10.1021/jz101536b} {\bibfield  {journal} {\bibinfo  {journal}
  {Journal of Physical Chemistry Letters}\ }\textbf {\bibinfo {volume} {2}},\
  \bibinfo {pages} {81} (\bibinfo {year} {2011})}\BibitemShut {NoStop}%
\bibitem [{\citenamefont {Kale}\ and\ \citenamefont
  {Herzfeld}(2012)}]{Kale2012/10.1002/anie.201203568}%
  \BibitemOpen
  \bibfield  {author} {\bibinfo {author} {\bibfnamefont {S.}~\bibnamefont
  {Kale}}\ and\ \bibinfo {author} {\bibfnamefont {J.}~\bibnamefont
  {Herzfeld}},\ }\href {\doibase 10.1002/anie.201203568} {\bibfield  {journal}
  {\bibinfo  {journal} {Angewandte Chemie International Edition}\ }\textbf
  {\bibinfo {volume} {51}},\ \bibinfo {pages} {11029} (\bibinfo {year}
  {2012})}\BibitemShut {NoStop}%
\bibitem [{\citenamefont {Hassanali}\ \emph {et~al.}(2013)\citenamefont
  {Hassanali}, \citenamefont {Giberti}, \citenamefont {Cuny}, \citenamefont
  {K{\"{u}}hne},\ and\ \citenamefont
  {Parrinello}}]{Hassanali2013/10.1073/pnas.1306642110}%
  \BibitemOpen
  \bibfield  {author} {\bibinfo {author} {\bibfnamefont {A.}~\bibnamefont
  {Hassanali}}, \bibinfo {author} {\bibfnamefont {F.}~\bibnamefont {Giberti}},
  \bibinfo {author} {\bibfnamefont {J.}~\bibnamefont {Cuny}}, \bibinfo {author}
  {\bibfnamefont {T.~D.}\ \bibnamefont {K{\"{u}}hne}}, \ and\ \bibinfo {author}
  {\bibfnamefont {M.}~\bibnamefont {Parrinello}},\ }\href {\doibase
  10.1073/pnas.1306642110} {\bibfield  {journal} {\bibinfo  {journal}
  {Proceedings of the National Academy of Sciences}\ }\textbf {\bibinfo
  {volume} {110}},\ \bibinfo {pages} {13723} (\bibinfo {year}
  {2013})}\BibitemShut {NoStop}%
\bibitem [{\citenamefont {Kulig}\ and\ \citenamefont
  {Agmon}(2013)}]{Kulig2013/10.1038/nchem.1503}%
  \BibitemOpen
  \bibfield  {author} {\bibinfo {author} {\bibfnamefont {W.}~\bibnamefont
  {Kulig}}\ and\ \bibinfo {author} {\bibfnamefont {N.}~\bibnamefont {Agmon}},\
  }\href {\doibase 10.1038/nchem.1503} {\bibfield  {journal} {\bibinfo
  {journal} {Nature Chemistry}\ }\textbf {\bibinfo {volume} {5}},\ \bibinfo
  {pages} {29} (\bibinfo {year} {2013})}\BibitemShut {NoStop}%
\bibitem [{\citenamefont {Baer}\ \emph
  {et~al.}(2014{\natexlab{a}})\citenamefont {Baer}, \citenamefont {Fulton},
  \citenamefont {Balasubramanian}, \citenamefont {Schenter},\ and\
  \citenamefont {Mundy}}]{Baer2014/10.1021/jp501091h}%
  \BibitemOpen
  \bibfield  {author} {\bibinfo {author} {\bibfnamefont {M.~D.}\ \bibnamefont
  {Baer}}, \bibinfo {author} {\bibfnamefont {J.~L.}\ \bibnamefont {Fulton}},
  \bibinfo {author} {\bibfnamefont {M.}~\bibnamefont {Balasubramanian}},
  \bibinfo {author} {\bibfnamefont {G.~K.}\ \bibnamefont {Schenter}}, \ and\
  \bibinfo {author} {\bibfnamefont {C.~J.}\ \bibnamefont {Mundy}},\ }\href
  {\doibase 10.1021/jp501091h} {\bibfield  {journal} {\bibinfo  {journal}
  {Journal of Physical Chemistry B}\ }\textbf {\bibinfo {volume} {118}},\
  \bibinfo {pages} {7211} (\bibinfo {year} {2014}{\natexlab{a}})}\BibitemShut
  {NoStop}%
\bibitem [{\citenamefont {Baer}\ \emph
  {et~al.}(2014{\natexlab{b}})\citenamefont {Baer}, \citenamefont {Kuo},
  \citenamefont {Tobias},\ and\ \citenamefont
  {Mundy}}]{Baer2014/10.1021/jp501854h}%
  \BibitemOpen
  \bibfield  {author} {\bibinfo {author} {\bibfnamefont {M.~D.}\ \bibnamefont
  {Baer}}, \bibinfo {author} {\bibfnamefont {I.-F.~W.}\ \bibnamefont {Kuo}},
  \bibinfo {author} {\bibfnamefont {D.~J.}\ \bibnamefont {Tobias}}, \ and\
  \bibinfo {author} {\bibfnamefont {C.~J.}\ \bibnamefont {Mundy}},\ }\href
  {\doibase 10.1021/jp501854h} {\bibfield  {journal} {\bibinfo  {journal} {The
  Journal of Physical Chemistry B}\ }\textbf {\bibinfo {volume} {118}},\
  \bibinfo {pages} {8364} (\bibinfo {year} {2014}{\natexlab{b}})}\BibitemShut
  {NoStop}%
\bibitem [{\citenamefont {Mancini}\ and\ \citenamefont
  {Bowman}(2015)}]{Mancini2015/10.1039/C4CP05685J}%
  \BibitemOpen
  \bibfield  {author} {\bibinfo {author} {\bibfnamefont {J.~S.}\ \bibnamefont
  {Mancini}}\ and\ \bibinfo {author} {\bibfnamefont {J.~M.}\ \bibnamefont
  {Bowman}},\ }\href {\doibase 10.1039/c4cp05685j} {\bibfield  {journal}
  {\bibinfo  {journal} {Phys. Chem. Chem. Phys.}\ }\textbf {\bibinfo {volume}
  {17}},\ \bibinfo {pages} {6222} (\bibinfo {year} {2015})}\BibitemShut
  {NoStop}%
\bibitem [{\citenamefont {Biswas}\ \emph {et~al.}(2017)\citenamefont {Biswas},
  \citenamefont {Carpenter}, \citenamefont {Fournier}, \citenamefont {Voth},\
  and\ \citenamefont {Tokmakoff}}]{Biswas2017/10.1063/1.4980121}%
  \BibitemOpen
  \bibfield  {author} {\bibinfo {author} {\bibfnamefont {R.}~\bibnamefont
  {Biswas}}, \bibinfo {author} {\bibfnamefont {W.}~\bibnamefont {Carpenter}},
  \bibinfo {author} {\bibfnamefont {J.~A.}\ \bibnamefont {Fournier}}, \bibinfo
  {author} {\bibfnamefont {G.~A.}\ \bibnamefont {Voth}}, \ and\ \bibinfo
  {author} {\bibfnamefont {A.}~\bibnamefont {Tokmakoff}},\ }\href {\doibase
  10.1063/1.4980121} {\bibfield  {journal} {\bibinfo  {journal} {Journal of
  Chemical Physics}\ }\textbf {\bibinfo {volume} {146}} (\bibinfo {year}
  {2017}),\ 10.1063/1.4980121}\BibitemShut {NoStop}%
\bibitem [{\citenamefont {Olesen}\ \emph {et~al.}(2011)\citenamefont {Olesen},
  \citenamefont {Guasco}, \citenamefont {Roscioli},\ and\ \citenamefont
  {Johnson}}]{Olesen2011/10.1016/j.cplett.2011.04.060}%
  \BibitemOpen
  \bibfield  {author} {\bibinfo {author} {\bibfnamefont {S.~G.}\ \bibnamefont
  {Olesen}}, \bibinfo {author} {\bibfnamefont {T.~L.}\ \bibnamefont {Guasco}},
  \bibinfo {author} {\bibfnamefont {J.~R.}\ \bibnamefont {Roscioli}}, \ and\
  \bibinfo {author} {\bibfnamefont {M.~A.}\ \bibnamefont {Johnson}},\ }\href
  {\doibase 10.1016/j.cplett.2011.04.060} {\bibfield  {journal} {\bibinfo
  {journal} {Chemical Physics Letters}\ }\textbf {\bibinfo {volume} {509}},\
  \bibinfo {pages} {89} (\bibinfo {year} {2011})}\BibitemShut {NoStop}%
\bibitem [{\citenamefont {Craig}\ \emph {et~al.}(2017)\citenamefont {Craig},
  \citenamefont {Menges}, \citenamefont {Duong}, \citenamefont {Denton},
  \citenamefont {Madison}, \citenamefont {McCoy},\ and\ \citenamefont
  {Johnson}}]{Craig2017/10.1073/pnas.1705089114}%
  \BibitemOpen
  \bibfield  {author} {\bibinfo {author} {\bibfnamefont {S.~M.}\ \bibnamefont
  {Craig}}, \bibinfo {author} {\bibfnamefont {F.~S.}\ \bibnamefont {Menges}},
  \bibinfo {author} {\bibfnamefont {C.~H.}\ \bibnamefont {Duong}}, \bibinfo
  {author} {\bibfnamefont {J.~K.}\ \bibnamefont {Denton}}, \bibinfo {author}
  {\bibfnamefont {L.~R.}\ \bibnamefont {Madison}}, \bibinfo {author}
  {\bibfnamefont {A.~B.}\ \bibnamefont {McCoy}}, \ and\ \bibinfo {author}
  {\bibfnamefont {M.~A.}\ \bibnamefont {Johnson}},\ }\href {\doibase
  10.1073/pnas.1705089114} {\bibfield  {journal} {\bibinfo  {journal}
  {Proceedings of the National Academy of Sciences}\ }\textbf {\bibinfo
  {volume} {114}},\ \bibinfo {pages} {E4706} (\bibinfo {year}
  {2017})}\BibitemShut {NoStop}%
\bibitem [{\citenamefont {Kim}\ \emph {et~al.}(2002)\citenamefont {Kim},
  \citenamefont {Schmitt}, \citenamefont {Gruetzmacher}, \citenamefont {Voth},\
  and\ \citenamefont {Scherer}}]{Kim2002/10.1063/1.1423327}%
  \BibitemOpen
  \bibfield  {author} {\bibinfo {author} {\bibfnamefont {J.}~\bibnamefont
  {Kim}}, \bibinfo {author} {\bibfnamefont {U.~W.}\ \bibnamefont {Schmitt}},
  \bibinfo {author} {\bibfnamefont {J.~A.}\ \bibnamefont {Gruetzmacher}},
  \bibinfo {author} {\bibfnamefont {G.~A.}\ \bibnamefont {Voth}}, \ and\
  \bibinfo {author} {\bibfnamefont {N.~E.}\ \bibnamefont {Scherer}},\ }\href
  {\doibase 10.1063/1.1423327} {\bibfield  {journal} {\bibinfo  {journal}
  {Journal of Chemical Physics}\ }\textbf {\bibinfo {volume} {116}},\ \bibinfo
  {pages} {737} (\bibinfo {year} {2002})}\BibitemShut {NoStop}%
\bibitem [{\citenamefont {Biswas}\ \emph {et~al.}(2016)\citenamefont {Biswas},
  \citenamefont {Carpenter}, \citenamefont {Voth},\ and\ \citenamefont
  {Tokmakoff}}]{Biswas2016/10.1063/1.4964723}%
  \BibitemOpen
  \bibfield  {author} {\bibinfo {author} {\bibfnamefont {R.}~\bibnamefont
  {Biswas}}, \bibinfo {author} {\bibfnamefont {W.}~\bibnamefont {Carpenter}},
  \bibinfo {author} {\bibfnamefont {G.~A.}\ \bibnamefont {Voth}}, \ and\
  \bibinfo {author} {\bibfnamefont {A.}~\bibnamefont {Tokmakoff}},\ }\href
  {\doibase 10.1063/1.4964723} {\bibfield  {journal} {\bibinfo  {journal}
  {Journal of Chemical Physics}\ }\textbf {\bibinfo {volume} {145}} (\bibinfo
  {year} {2016}),\ 10.1063/1.4964723}\BibitemShut {NoStop}%
\bibitem [{\citenamefont {Lin}\ and\ \citenamefont
  {Paesani}(2015)}]{Lin2015/10.1021/jp509791n}%
  \BibitemOpen
  \bibfield  {author} {\bibinfo {author} {\bibfnamefont {W.}~\bibnamefont
  {Lin}}\ and\ \bibinfo {author} {\bibfnamefont {F.}~\bibnamefont {Paesani}},\
  }\href {\doibase 10.1021/jp509791n} {\bibfield  {journal} {\bibinfo
  {journal} {The Journal of Physical Chemistry A}\ }\textbf {\bibinfo {volume}
  {119}},\ \bibinfo {pages} {4450} (\bibinfo {year} {2015})}\BibitemShut
  {NoStop}%
\bibitem [{\citenamefont {Marsalek}\ and\ \citenamefont
  {Markland}(2017)}]{Marsalek2017/10.1021/acs.jpclett.7b00391}%
  \BibitemOpen
  \bibfield  {author} {\bibinfo {author} {\bibfnamefont {O.}~\bibnamefont
  {Marsalek}}\ and\ \bibinfo {author} {\bibfnamefont {T.~E.}\ \bibnamefont
  {Markland}},\ }\href {\doibase 10.1021/acs.jpclett.7b00391} {\bibfield
  {journal} {\bibinfo  {journal} {The Journal of Physical Chemistry Letters}\
  }\textbf {\bibinfo {volume} {8}},\ \bibinfo {pages} {1545} (\bibinfo {year}
  {2017})}\BibitemShut {NoStop}%
\bibitem [{\citenamefont {Asmis}\ \emph {et~al.}(2003)\citenamefont {Asmis},
  \citenamefont {Pivonka}, \citenamefont {Santambrogio}, \citenamefont
  {Br{\"{u}}mmer}, \citenamefont {Kaposta}, \citenamefont {Neumark},\ and\
  \citenamefont {W{\"{o}}ste}}]{Asmis2003/10.1126/science.1081634}%
  \BibitemOpen
  \bibfield  {author} {\bibinfo {author} {\bibfnamefont {K.~R.}\ \bibnamefont
  {Asmis}}, \bibinfo {author} {\bibfnamefont {N.~L.}\ \bibnamefont {Pivonka}},
  \bibinfo {author} {\bibfnamefont {G.}~\bibnamefont {Santambrogio}}, \bibinfo
  {author} {\bibfnamefont {M.}~\bibnamefont {Br{\"{u}}mmer}}, \bibinfo {author}
  {\bibfnamefont {C.}~\bibnamefont {Kaposta}}, \bibinfo {author} {\bibfnamefont
  {D.~M.}\ \bibnamefont {Neumark}}, \ and\ \bibinfo {author} {\bibfnamefont
  {L.}~\bibnamefont {W{\"{o}}ste}},\ }\href {\doibase 10.1126/science.1081634}
  {\bibfield  {journal} {\bibinfo  {journal} {Science}\ }\textbf {\bibinfo
  {volume} {299}},\ \bibinfo {pages} {1375} (\bibinfo {year}
  {2003})}\BibitemShut {NoStop}%
\bibitem [{\citenamefont {McQuarrie}(2000)}]{McQuarrie}%
  \BibitemOpen
  \bibfield  {author} {\bibinfo {author} {\bibfnamefont {D.~A.}\ \bibnamefont
  {McQuarrie}},\ }\href
  {https://www.amazon.com/Statistical-Mechanics-Donald-Allan-McQuarrie/dp/1891389157?SubscriptionId=0JYN1NVW651KCA56C102&tag=techkie-20&linkCode=xm2&camp=2025&creative=165953&creativeASIN=1891389157}
  {\emph {\bibinfo {title} {Statistical Mechanics}}}\ (\bibinfo  {publisher}
  {University Science Books},\ \bibinfo {year} {2000})\BibitemShut {NoStop}%
\bibitem [{\citenamefont {Lapid}\ \emph {et~al.}(2005)\citenamefont {Lapid},
  \citenamefont {Agmon}, \citenamefont {Petersen},\ and\ \citenamefont
  {Voth}}]{Lapid2005/10.1063/1.1814973}%
  \BibitemOpen
  \bibfield  {author} {\bibinfo {author} {\bibfnamefont {H.}~\bibnamefont
  {Lapid}}, \bibinfo {author} {\bibfnamefont {N.}~\bibnamefont {Agmon}},
  \bibinfo {author} {\bibfnamefont {M.~K.}\ \bibnamefont {Petersen}}, \ and\
  \bibinfo {author} {\bibfnamefont {G.~A.}\ \bibnamefont {Voth}},\ }\href
  {\doibase 10.1063/1.1814973} {\bibfield  {journal} {\bibinfo  {journal}
  {Journal of Chemical Physics}\ }\textbf {\bibinfo {volume} {122}},\ \bibinfo
  {pages} {014506} (\bibinfo {year} {2005})}\BibitemShut {NoStop}%
\bibitem [{\citenamefont {Ohno}\ \emph {et~al.}(2005)\citenamefont {Ohno},
  \citenamefont {Okimura}, \citenamefont {Akai},\ and\ \citenamefont
  {Katsumoto}}]{Ohno2005/10.1039/b506641g}%
  \BibitemOpen
  \bibfield  {author} {\bibinfo {author} {\bibfnamefont {K.}~\bibnamefont
  {Ohno}}, \bibinfo {author} {\bibfnamefont {M.}~\bibnamefont {Okimura}},
  \bibinfo {author} {\bibfnamefont {N.}~\bibnamefont {Akai}}, \ and\ \bibinfo
  {author} {\bibfnamefont {Y.}~\bibnamefont {Katsumoto}},\ }\href {\doibase
  10.1039/b506641g} {\bibfield  {journal} {\bibinfo  {journal} {Physical
  Chemistry Chemical Physics}\ }\textbf {\bibinfo {volume} {7}},\ \bibinfo
  {pages} {3005} (\bibinfo {year} {2005})}\BibitemShut {NoStop}%
\bibitem [{\citenamefont {Sun}(2009)}]{Sun2009/10.1016/j.vibspec.2009.05.002}%
  \BibitemOpen
  \bibfield  {author} {\bibinfo {author} {\bibfnamefont {Q.}~\bibnamefont
  {Sun}},\ }\href {\doibase 10.1016/j.vibspec.2009.05.002} {\bibfield
  {journal} {\bibinfo  {journal} {Vibrational Spectroscopy}\ }\textbf {\bibinfo
  {volume} {51}},\ \bibinfo {pages} {213} (\bibinfo {year} {2009})}\BibitemShut
  {NoStop}%
\bibitem [{\citenamefont {V{\'{a}}cha}\ \emph {et~al.}(2012)\citenamefont
  {V{\'{a}}cha}, \citenamefont {Marsalek}, \citenamefont {Willard},
  \citenamefont {Bonthuis}, \citenamefont {Netz},\ and\ \citenamefont
  {Jungwirth}}]{Vacha2012/10.1021/jz2014852}%
  \BibitemOpen
  \bibfield  {author} {\bibinfo {author} {\bibfnamefont {R.}~\bibnamefont
  {V{\'{a}}cha}}, \bibinfo {author} {\bibfnamefont {O.}~\bibnamefont
  {Marsalek}}, \bibinfo {author} {\bibfnamefont {A.~P.}\ \bibnamefont
  {Willard}}, \bibinfo {author} {\bibfnamefont {D.~J.}\ \bibnamefont
  {Bonthuis}}, \bibinfo {author} {\bibfnamefont {R.~R.}\ \bibnamefont {Netz}},
  \ and\ \bibinfo {author} {\bibfnamefont {P.}~\bibnamefont {Jungwirth}},\
  }\href {\doibase 10.1021/jz2014852} {\bibfield  {journal} {\bibinfo
  {journal} {The Journal of Physical Chemistry Letters}\ }\textbf {\bibinfo
  {volume} {3}},\ \bibinfo {pages} {107} (\bibinfo {year} {2012})}\BibitemShut
  {NoStop}%
\bibitem [{\citenamefont {Tainter}\ \emph {et~al.}(2013)\citenamefont
  {Tainter}, \citenamefont {Ni}, \citenamefont {Shi},\ and\ \citenamefont
  {Skinner}}]{Tainter2013/10.1021/jz301780k}%
  \BibitemOpen
  \bibfield  {author} {\bibinfo {author} {\bibfnamefont {C.~J.}\ \bibnamefont
  {Tainter}}, \bibinfo {author} {\bibfnamefont {Y.}~\bibnamefont {Ni}},
  \bibinfo {author} {\bibfnamefont {L.}~\bibnamefont {Shi}}, \ and\ \bibinfo
  {author} {\bibfnamefont {J.~L.}\ \bibnamefont {Skinner}},\ }\href {\doibase
  10.1021/jz301780k} {\bibfield  {journal} {\bibinfo  {journal} {The Journal of
  Physical Chemistry Letters}\ }\textbf {\bibinfo {volume} {4}},\ \bibinfo
  {pages} {12} (\bibinfo {year} {2013})}\BibitemShut {NoStop}%
\bibitem [{\citenamefont {Berkelbach}, \citenamefont {Lee},\ and\ \citenamefont
  {Tuckerman}(2009)}]{Berkelbach2009/10.1103/PhysRevLett.103.238302}%
  \BibitemOpen
  \bibfield  {author} {\bibinfo {author} {\bibfnamefont {T.~C.}\ \bibnamefont
  {Berkelbach}}, \bibinfo {author} {\bibfnamefont {H.~S.}\ \bibnamefont {Lee}},
  \ and\ \bibinfo {author} {\bibfnamefont {M.~E.}\ \bibnamefont {Tuckerman}},\
  }\href {\doibase 10.1103/PhysRevLett.103.238302} {\bibfield  {journal}
  {\bibinfo  {journal} {Physical Review Letters}\ }\textbf {\bibinfo {volume}
  {103}},\ \bibinfo {pages} {1} (\bibinfo {year} {2009})}\BibitemShut {NoStop}%
\bibitem [{\citenamefont {Tuckerman}, \citenamefont {Chandra},\ and\
  \citenamefont {Marx}(2010)}]{Tuckerman2010/10.1063/1.3474625}%
  \BibitemOpen
  \bibfield  {author} {\bibinfo {author} {\bibfnamefont {M.~E.}\ \bibnamefont
  {Tuckerman}}, \bibinfo {author} {\bibfnamefont {A.}~\bibnamefont {Chandra}},
  \ and\ \bibinfo {author} {\bibfnamefont {D.}~\bibnamefont {Marx}},\ }\href
  {\doibase 10.1063/1.3474625} {\bibfield  {journal} {\bibinfo  {journal} {The
  Journal of Chemical Physics}\ }\textbf {\bibinfo {volume} {133}},\ \bibinfo
  {pages} {124108} (\bibinfo {year} {2010})}\BibitemShut {NoStop}%
\bibitem [{\citenamefont {Marx}, \citenamefont {Tuckerman},\ and\ \citenamefont
  {Parrinello}(2000)}]{Marx2000/10.1088/0953-8984/12/8A/317}%
  \BibitemOpen
  \bibfield  {author} {\bibinfo {author} {\bibfnamefont {D.}~\bibnamefont
  {Marx}}, \bibinfo {author} {\bibfnamefont {M.~E.}\ \bibnamefont {Tuckerman}},
  \ and\ \bibinfo {author} {\bibfnamefont {M.}~\bibnamefont {Parrinello}},\
  }\href {\doibase 10.1088/0953-8984/12/8A/317} {\bibfield  {journal} {\bibinfo
   {journal} {Journal of Physics: Condensed Matter}\ }\textbf {\bibinfo
  {volume} {12}},\ \bibinfo {pages} {A153} (\bibinfo {year}
  {2000})}\BibitemShut {NoStop}%
\bibitem [{\citenamefont {Dahms}\ \emph {et~al.}(2017)\citenamefont {Dahms},
  \citenamefont {Fingerhut}, \citenamefont {Nibbering}, \citenamefont {Pines},\
  and\ \citenamefont {Elsaesser}}]{Dahms2017/10.1126/science.aan5144}%
  \BibitemOpen
  \bibfield  {author} {\bibinfo {author} {\bibfnamefont {F.}~\bibnamefont
  {Dahms}}, \bibinfo {author} {\bibfnamefont {B.~P.}\ \bibnamefont
  {Fingerhut}}, \bibinfo {author} {\bibfnamefont {E.~T.~J.}\ \bibnamefont
  {Nibbering}}, \bibinfo {author} {\bibfnamefont {E.}~\bibnamefont {Pines}}, \
  and\ \bibinfo {author} {\bibfnamefont {T.}~\bibnamefont {Elsaesser}},\ }\href
  {\doibase 10.1126/science.aan5144} {\bibfield  {journal} {\bibinfo  {journal}
  {Science}\ }\textbf {\bibinfo {volume} {357}},\ \bibinfo {pages} {491}
  (\bibinfo {year} {2017})}\BibitemShut {NoStop}%
\bibitem [{\citenamefont {Marsalek}\ and\ \citenamefont
  {Markland}(2016)}]{Marsalek2016/10.1063/1.4941093}%
  \BibitemOpen
  \bibfield  {author} {\bibinfo {author} {\bibfnamefont {O.}~\bibnamefont
  {Marsalek}}\ and\ \bibinfo {author} {\bibfnamefont {T.~E.}\ \bibnamefont
  {Markland}},\ }\href {\doibase 10.1063/1.4941093} {\bibfield  {journal}
  {\bibinfo  {journal} {The Journal of Chemical Physics}\ }\textbf {\bibinfo
  {volume} {144}},\ \bibinfo {pages} {054112} (\bibinfo {year} {2016})},\
  \Eprint {http://arxiv.org/abs/1512.00473} {arXiv:1512.00473} \BibitemShut
  {NoStop}%
\bibitem [{\citenamefont {Green}\ and\ \citenamefont
  {Perry}(2007)}]{perry-handbook}%
  \BibitemOpen
  \bibfield  {author} {\bibinfo {author} {\bibfnamefont {D.}~\bibnamefont
  {Green}}\ and\ \bibinfo {author} {\bibfnamefont {R.}~\bibnamefont {Perry}},\
  }\href
  {https://www.amazon.com/Perrys-Chemical-Engineers-Handbook-Eighth-ebook/dp/B00LCRO9O8?SubscriptionId=0JYN1NVW651KCA56C102&tag=techkie-20&linkCode=xm2&camp=2025&creative=165953&creativeASIN=B00LCRO9O8}
  {\emph {\bibinfo {title} {Perry's Chemical Engineers' Handbook, Eighth
  Edition (Chemical Engineers Handbook)}}}\ (\bibinfo  {publisher} {McGraw-Hill
  Education},\ \bibinfo {year} {2007})\BibitemShut {NoStop}%
\bibitem [{\citenamefont {Shi}\ \emph {et~al.}(2013)\citenamefont {Shi},
  \citenamefont {Xia}, \citenamefont {Zhang}, \citenamefont {Best},
  \citenamefont {Wu}, \citenamefont {Ponder},\ and\ \citenamefont
  {Ren}}]{Shi2013/10.1021/ct4003702}%
  \BibitemOpen
  \bibfield  {author} {\bibinfo {author} {\bibfnamefont {Y.}~\bibnamefont
  {Shi}}, \bibinfo {author} {\bibfnamefont {Z.}~\bibnamefont {Xia}}, \bibinfo
  {author} {\bibfnamefont {J.}~\bibnamefont {Zhang}}, \bibinfo {author}
  {\bibfnamefont {R.}~\bibnamefont {Best}}, \bibinfo {author} {\bibfnamefont
  {C.}~\bibnamefont {Wu}}, \bibinfo {author} {\bibfnamefont {J.~W.}\
  \bibnamefont {Ponder}}, \ and\ \bibinfo {author} {\bibfnamefont
  {P.}~\bibnamefont {Ren}},\ }\href {\doibase 10.1021/ct4003702} {\bibfield
  {journal} {\bibinfo  {journal} {Journal of Chemical Theory and Computation}\
  }\textbf {\bibinfo {volume} {9}},\ \bibinfo {pages} {4046} (\bibinfo {year}
  {2013})}\BibitemShut {NoStop}%
\bibitem [{\citenamefont {Ceriotti}, \citenamefont {More},\ and\ \citenamefont
  {Manolopoulos}(2013)}]{Ceriotti2013/10.1016/j.cpc.2013.10.027}%
  \BibitemOpen
  \bibfield  {author} {\bibinfo {author} {\bibfnamefont {M.}~\bibnamefont
  {Ceriotti}}, \bibinfo {author} {\bibfnamefont {J.}~\bibnamefont {More}}, \
  and\ \bibinfo {author} {\bibfnamefont {D.~E.}\ \bibnamefont {Manolopoulos}},\
  }\href {\doibase 10.1016/j.cpc.2013.10.027} {\bibfield  {journal} {\bibinfo
  {journal} {Computer Physics Communications}\ }\textbf {\bibinfo {volume}
  {185}},\ \bibinfo {pages} {1019} (\bibinfo {year} {2013})}\BibitemShut
  {NoStop}%
\bibitem [{\citenamefont {Tuckerman}, \citenamefont {Berne},\ and\
  \citenamefont {Martyna}(1992)}]{Tuckerman1992/10.1063/1.463137}%
  \BibitemOpen
  \bibfield  {author} {\bibinfo {author} {\bibfnamefont {M.}~\bibnamefont
  {Tuckerman}}, \bibinfo {author} {\bibfnamefont {B.~J.}\ \bibnamefont
  {Berne}}, \ and\ \bibinfo {author} {\bibfnamefont {G.~J.}\ \bibnamefont
  {Martyna}},\ }\href {\doibase 10.1063/1.463137} {\bibfield  {journal}
  {\bibinfo  {journal} {The Journal of Chemical Physics}\ }\textbf {\bibinfo
  {volume} {97}},\ \bibinfo {pages} {1990} (\bibinfo {year}
  {1992})}\BibitemShut {NoStop}%
\bibitem [{\citenamefont {Bussi}, \citenamefont {Donadio},\ and\ \citenamefont
  {Parrinello}(2007)}]{Bussi2007/10.1063/1.2408420}%
  \BibitemOpen
  \bibfield  {author} {\bibinfo {author} {\bibfnamefont {G.}~\bibnamefont
  {Bussi}}, \bibinfo {author} {\bibfnamefont {D.}~\bibnamefont {Donadio}}, \
  and\ \bibinfo {author} {\bibfnamefont {M.}~\bibnamefont {Parrinello}},\
  }\href {\doibase 10.1063/1.2408420} {\bibfield  {journal} {\bibinfo
  {journal} {The Journal of chemical physics}\ }\textbf {\bibinfo {volume}
  {126}},\ \bibinfo {pages} {014101} (\bibinfo {year} {2007})}\BibitemShut
  {NoStop}%
\bibitem [{\citenamefont {Markland}\ and\ \citenamefont
  {Manolopoulos}(2008{\natexlab{a}})}]{Markland2008/10.1063/1.2953308}%
  \BibitemOpen
  \bibfield  {author} {\bibinfo {author} {\bibfnamefont {T.~E.}\ \bibnamefont
  {Markland}}\ and\ \bibinfo {author} {\bibfnamefont {D.~E.}\ \bibnamefont
  {Manolopoulos}},\ }\href {\doibase 10.1063/1.2953308} {\bibfield  {journal}
  {\bibinfo  {journal} {The Journal of chemical physics}\ }\textbf {\bibinfo
  {volume} {129}},\ \bibinfo {pages} {024105} (\bibinfo {year}
  {2008}{\natexlab{a}})}\BibitemShut {NoStop}%
\bibitem [{\citenamefont {Markland}\ and\ \citenamefont
  {Manolopoulos}(2008{\natexlab{b}})}]{Markland2008/10.1016/j.cplett.2008.09.019}%
  \BibitemOpen
  \bibfield  {author} {\bibinfo {author} {\bibfnamefont {T.~E.}\ \bibnamefont
  {Markland}}\ and\ \bibinfo {author} {\bibfnamefont {D.~E.}\ \bibnamefont
  {Manolopoulos}},\ }\href {\doibase 10.1016/j.cplett.2008.09.019} {\bibfield
  {journal} {\bibinfo  {journal} {Chemical Physics Letters}\ }\textbf {\bibinfo
  {volume} {464}},\ \bibinfo {pages} {256} (\bibinfo {year}
  {2008}{\natexlab{b}})}\BibitemShut {NoStop}%
\bibitem [{\citenamefont {Craig}\ and\ \citenamefont
  {Manolopoulos}(2004)}]{Craig2004/10.1063/1.1777575}%
  \BibitemOpen
  \bibfield  {author} {\bibinfo {author} {\bibfnamefont {I.~R.}\ \bibnamefont
  {Craig}}\ and\ \bibinfo {author} {\bibfnamefont {D.~E.}\ \bibnamefont
  {Manolopoulos}},\ }\href {\doibase 10.1063/1.1777575} {\bibfield  {journal}
  {\bibinfo  {journal} {The Journal of Chemical Physics}\ }\textbf {\bibinfo
  {volume} {121}},\ \bibinfo {pages} {3368} (\bibinfo {year}
  {2004})}\BibitemShut {NoStop}%
\bibitem [{\citenamefont {Habershon}\ \emph {et~al.}(2013)\citenamefont
  {Habershon}, \citenamefont {Manolopoulos}, \citenamefont {Markland},\ and\
  \citenamefont
  {Miller}}]{Habershon2013/10.1146/annurev-physchem-040412-110122}%
  \BibitemOpen
  \bibfield  {author} {\bibinfo {author} {\bibfnamefont {S.}~\bibnamefont
  {Habershon}}, \bibinfo {author} {\bibfnamefont {D.~E.}\ \bibnamefont
  {Manolopoulos}}, \bibinfo {author} {\bibfnamefont {T.~E.}\ \bibnamefont
  {Markland}}, \ and\ \bibinfo {author} {\bibfnamefont {T.~F.}\ \bibnamefont
  {Miller}},\ }\href {\doibase 10.1146/annurev-physchem-040412-110122}
  {\bibfield  {journal} {\bibinfo  {journal} {Annual review of physical
  chemistry}\ }\textbf {\bibinfo {volume} {64}},\ \bibinfo {pages} {387}
  (\bibinfo {year} {2013})}\BibitemShut {NoStop}%
\bibitem [{\citenamefont {Rossi}, \citenamefont {Ceriotti},\ and\ \citenamefont
  {Manolopoulos}(2014)}]{Rossi2014/10.1063/1.4883861}%
  \BibitemOpen
  \bibfield  {author} {\bibinfo {author} {\bibfnamefont {M.}~\bibnamefont
  {Rossi}}, \bibinfo {author} {\bibfnamefont {M.}~\bibnamefont {Ceriotti}}, \
  and\ \bibinfo {author} {\bibfnamefont {D.~E.}\ \bibnamefont {Manolopoulos}},\
  }\href {\doibase 10.1063/1.4883861} {\bibfield  {journal} {\bibinfo
  {journal} {Journal of Chemical Physics}\ }\textbf {\bibinfo {volume} {140}},\
  \bibinfo {pages} {234116} (\bibinfo {year} {2014})},\ \Eprint
  {http://arxiv.org/abs/1406.1074v1} {arXiv:1406.1074v1} \BibitemShut {NoStop}%
\bibitem [{\citenamefont {Ceriotti}\ \emph {et~al.}(2010)\citenamefont
  {Ceriotti}, \citenamefont {Parrinello}, \citenamefont {Markland},\ and\
  \citenamefont {Manolopoulos}}]{Ceriotti2010/10.1063/1.3489925}%
  \BibitemOpen
  \bibfield  {author} {\bibinfo {author} {\bibfnamefont {M.}~\bibnamefont
  {Ceriotti}}, \bibinfo {author} {\bibfnamefont {M.}~\bibnamefont
  {Parrinello}}, \bibinfo {author} {\bibfnamefont {T.~E.}\ \bibnamefont
  {Markland}}, \ and\ \bibinfo {author} {\bibfnamefont {D.~E.}\ \bibnamefont
  {Manolopoulos}},\ }\href {\doibase 10.1063/1.3489925} {\bibfield  {journal}
  {\bibinfo  {journal} {The Journal of chemical physics}\ }\textbf {\bibinfo
  {volume} {133}},\ \bibinfo {pages} {124104} (\bibinfo {year}
  {2010})}\BibitemShut {NoStop}%
\bibitem [{\citenamefont {VandeVondele}\ \emph {et~al.}(2005)\citenamefont
  {VandeVondele}, \citenamefont {Krack}, \citenamefont {Mohamed}, \citenamefont
  {Parrinello}, \citenamefont {Chassaing},\ and\ \citenamefont
  {Hutter}}]{Vandevondele2005/10.1016/j.cpc.2004.12.014}%
  \BibitemOpen
  \bibfield  {author} {\bibinfo {author} {\bibfnamefont {J.}~\bibnamefont
  {VandeVondele}}, \bibinfo {author} {\bibfnamefont {M.}~\bibnamefont {Krack}},
  \bibinfo {author} {\bibfnamefont {F.}~\bibnamefont {Mohamed}}, \bibinfo
  {author} {\bibfnamefont {M.}~\bibnamefont {Parrinello}}, \bibinfo {author}
  {\bibfnamefont {T.}~\bibnamefont {Chassaing}}, \ and\ \bibinfo {author}
  {\bibfnamefont {J.}~\bibnamefont {Hutter}},\ }\href {\doibase
  10.1016/j.cpc.2004.12.014} {\bibfield  {journal} {\bibinfo  {journal}
  {Computer Physics Communications}\ }\textbf {\bibinfo {volume} {167}},\
  \bibinfo {pages} {103} (\bibinfo {year} {2005})}\BibitemShut {NoStop}%
\bibitem [{\citenamefont {Hutter}\ \emph {et~al.}(2014)\citenamefont {Hutter},
  \citenamefont {Iannuzzi}, \citenamefont {Schiffmann},\ and\ \citenamefont
  {VandeVondele}}]{Hutter2014/10.1002/wcms.1159}%
  \BibitemOpen
  \bibfield  {author} {\bibinfo {author} {\bibfnamefont {J.}~\bibnamefont
  {Hutter}}, \bibinfo {author} {\bibfnamefont {M.}~\bibnamefont {Iannuzzi}},
  \bibinfo {author} {\bibfnamefont {F.}~\bibnamefont {Schiffmann}}, \ and\
  \bibinfo {author} {\bibfnamefont {J.}~\bibnamefont {VandeVondele}},\ }\href
  {\doibase 10.1002/wcms.1159} {\bibfield  {journal} {\bibinfo  {journal}
  {Wiley Interdisciplinary Reviews: Computational Molecular Science}\ }\textbf
  {\bibinfo {volume} {4}},\ \bibinfo {pages} {15} (\bibinfo {year}
  {2014})}\BibitemShut {NoStop}%
\bibitem [{\citenamefont {Perdew}, \citenamefont {Burke},\ and\ \citenamefont
  {Ernzerhof}(1996)}]{Perdew1996/10.1103/PhysRevLett.77.3865}%
  \BibitemOpen
  \bibfield  {author} {\bibinfo {author} {\bibfnamefont {J.~P.}\ \bibnamefont
  {Perdew}}, \bibinfo {author} {\bibfnamefont {K.}~\bibnamefont {Burke}}, \
  and\ \bibinfo {author} {\bibfnamefont {M.}~\bibnamefont {Ernzerhof}},\ }\href
  {\doibase 10.1103/PhysRevLett.77.3865} {\bibfield  {journal} {\bibinfo
  {journal} {Physical Review Letters}\ }\textbf {\bibinfo {volume} {77}},\
  \bibinfo {pages} {3865} (\bibinfo {year} {1996})}\BibitemShut {NoStop}%
\bibitem [{\citenamefont {Zhang}\ and\ \citenamefont
  {Yang}(1998)}]{Zhang1998/10.1103/PhysRevLett.80.890}%
  \BibitemOpen
  \bibfield  {author} {\bibinfo {author} {\bibfnamefont {Y.}~\bibnamefont
  {Zhang}}\ and\ \bibinfo {author} {\bibfnamefont {W.}~\bibnamefont {Yang}},\
  }\href {\doibase 10.1103/PhysRevLett.80.890} {\bibfield  {journal} {\bibinfo
  {journal} {Physical Review Letters}\ }\textbf {\bibinfo {volume} {80}},\
  \bibinfo {pages} {890} (\bibinfo {year} {1998})}\BibitemShut {NoStop}%
\bibitem [{\citenamefont {Adamo}\ and\ \citenamefont
  {Barone}(1999)}]{Adamo1999/10.1063/1.478522}%
  \BibitemOpen
  \bibfield  {author} {\bibinfo {author} {\bibfnamefont {C.}~\bibnamefont
  {Adamo}}\ and\ \bibinfo {author} {\bibfnamefont {V.}~\bibnamefont {Barone}},\
  }\href {\doibase 10.1063/1.478522} {\bibfield  {journal} {\bibinfo  {journal}
  {The Journal of Chemical Physics}\ }\textbf {\bibinfo {volume} {110}},\
  \bibinfo {pages} {6158} (\bibinfo {year} {1999})}\BibitemShut {NoStop}%
\bibitem [{\citenamefont {Goerigk}\ and\ \citenamefont
  {Grimme}(2011)}]{Goerigk2011/10.1039/c0cp02984j}%
  \BibitemOpen
  \bibfield  {author} {\bibinfo {author} {\bibfnamefont {L.}~\bibnamefont
  {Goerigk}}\ and\ \bibinfo {author} {\bibfnamefont {S.}~\bibnamefont
  {Grimme}},\ }\href {\doibase 10.1039/c0cp02984j} {\bibfield  {journal}
  {\bibinfo  {journal} {Physical Chemistry Chemical Physics}\ }\textbf
  {\bibinfo {volume} {13}},\ \bibinfo {pages} {6670} (\bibinfo {year}
  {2011})}\BibitemShut {NoStop}%
\bibitem [{\citenamefont {Grimme}\ \emph {et~al.}(2010)\citenamefont {Grimme},
  \citenamefont {Antony}, \citenamefont {Ehrlich},\ and\ \citenamefont
  {Krieg}}]{Grimme2010/10.1063/1.3382344}%
  \BibitemOpen
  \bibfield  {author} {\bibinfo {author} {\bibfnamefont {S.}~\bibnamefont
  {Grimme}}, \bibinfo {author} {\bibfnamefont {J.}~\bibnamefont {Antony}},
  \bibinfo {author} {\bibfnamefont {S.}~\bibnamefont {Ehrlich}}, \ and\
  \bibinfo {author} {\bibfnamefont {H.}~\bibnamefont {Krieg}},\ }\href
  {\doibase 10.1063/1.3382344} {\bibfield  {journal} {\bibinfo  {journal} {The
  Journal of chemical physics}\ }\textbf {\bibinfo {volume} {132}},\ \bibinfo
  {pages} {154104} (\bibinfo {year} {2010})}\BibitemShut {NoStop}%
\bibitem [{\citenamefont {Goedecker}, \citenamefont {Teter},\ and\
  \citenamefont {Hutter}(1996)}]{Goedecker1996/10.1103/PhysRevB.54.1703}%
  \BibitemOpen
  \bibfield  {author} {\bibinfo {author} {\bibfnamefont {S.}~\bibnamefont
  {Goedecker}}, \bibinfo {author} {\bibfnamefont {M.}~\bibnamefont {Teter}}, \
  and\ \bibinfo {author} {\bibfnamefont {J.}~\bibnamefont {Hutter}},\ }\href
  {\doibase 10.1103/PhysRevB.54.1703} {\bibfield  {journal} {\bibinfo
  {journal} {Physical Review B}\ }\textbf {\bibinfo {volume} {54}},\ \bibinfo
  {pages} {1703} (\bibinfo {year} {1996})}\BibitemShut {NoStop}%
\bibitem [{\citenamefont {Lippert}, \citenamefont {Hutter},\ and\ \citenamefont
  {Parrinello}(1997)}]{Lippert1997/10.1080/002689797170220}%
  \BibitemOpen
  \bibfield  {author} {\bibinfo {author} {\bibfnamefont {G.}~\bibnamefont
  {Lippert}}, \bibinfo {author} {\bibfnamefont {J.}~\bibnamefont {Hutter}}, \
  and\ \bibinfo {author} {\bibfnamefont {M.}~\bibnamefont {Parrinello}},\
  }\href {\doibase 10.1080/002689797170220} {\bibfield  {journal} {\bibinfo
  {journal} {Molecular Physics}\ }\textbf {\bibinfo {volume} {92}},\ \bibinfo
  {pages} {477} (\bibinfo {year} {1997})}\BibitemShut {NoStop}%
\bibitem [{\citenamefont {Guidon}, \citenamefont {Hutter},\ and\ \citenamefont
  {VandeVondele}(2009)}]{Guidon2009/10.1021/ct900494g}%
  \BibitemOpen
  \bibfield  {author} {\bibinfo {author} {\bibfnamefont {M.}~\bibnamefont
  {Guidon}}, \bibinfo {author} {\bibfnamefont {J.}~\bibnamefont {Hutter}}, \
  and\ \bibinfo {author} {\bibfnamefont {J.}~\bibnamefont {VandeVondele}},\
  }\href {\doibase 10.1021/ct900494g} {\bibfield  {journal} {\bibinfo
  {journal} {Journal of Chemical Theory and Computation}\ }\textbf {\bibinfo
  {volume} {5}},\ \bibinfo {pages} {3010} (\bibinfo {year} {2009})}\BibitemShut
  {NoStop}%
\bibitem [{\citenamefont {Guidon}, \citenamefont {Hutter},\ and\ \citenamefont
  {VandeVondele}(2010)}]{Guidon2010/10.1021/ct1002225}%
  \BibitemOpen
  \bibfield  {author} {\bibinfo {author} {\bibfnamefont {M.}~\bibnamefont
  {Guidon}}, \bibinfo {author} {\bibfnamefont {J.}~\bibnamefont {Hutter}}, \
  and\ \bibinfo {author} {\bibfnamefont {J.}~\bibnamefont {VandeVondele}},\
  }\href {\doibase 10.1021/ct1002225} {\bibfield  {journal} {\bibinfo
  {journal} {Journal of Chemical Theory and Computation}\ }\textbf {\bibinfo
  {volume} {6}},\ \bibinfo {pages} {2348} (\bibinfo {year} {2010})}\BibitemShut
  {NoStop}%
\bibitem [{\citenamefont {VandeVondele}\ and\ \citenamefont
  {Hutter}(2003)}]{VandeVondele2003/10.1063/1.1543154}%
  \BibitemOpen
  \bibfield  {author} {\bibinfo {author} {\bibfnamefont {J.}~\bibnamefont
  {VandeVondele}}\ and\ \bibinfo {author} {\bibfnamefont {J.}~\bibnamefont
  {Hutter}},\ }\href {\doibase 10.1063/1.1543154} {\bibfield  {journal}
  {\bibinfo  {journal} {The Journal of Chemical Physics}\ }\textbf {\bibinfo
  {volume} {118}},\ \bibinfo {pages} {4365} (\bibinfo {year}
  {2003})}\BibitemShut {NoStop}%
\bibitem [{\citenamefont {Kolafa}(2004)}]{Kolafa2004/10.1002/jcc.10385}%
  \BibitemOpen
  \bibfield  {author} {\bibinfo {author} {\bibfnamefont {J.}~\bibnamefont
  {Kolafa}},\ }\href {\doibase 10.1002/jcc.10385} {\bibfield  {journal}
  {\bibinfo  {journal} {Journal of computational chemistry}\ }\textbf {\bibinfo
  {volume} {25}},\ \bibinfo {pages} {335} (\bibinfo {year} {2004})}\BibitemShut
  {NoStop}%
\bibitem [{\citenamefont {K{\"{u}}hne}\ \emph {et~al.}(2007)\citenamefont
  {K{\"{u}}hne}, \citenamefont {Krack}, \citenamefont {Mohamed},\ and\
  \citenamefont {Parrinello}}]{Kuhne2007/10.1103/PhysRevLett.98.066401}%
  \BibitemOpen
  \bibfield  {author} {\bibinfo {author} {\bibfnamefont {T.}~\bibnamefont
  {K{\"{u}}hne}}, \bibinfo {author} {\bibfnamefont {M.}~\bibnamefont {Krack}},
  \bibinfo {author} {\bibfnamefont {F.}~\bibnamefont {Mohamed}}, \ and\
  \bibinfo {author} {\bibfnamefont {M.}~\bibnamefont {Parrinello}},\ }\href
  {\doibase 10.1103/PhysRevLett.98.066401} {\bibfield  {journal} {\bibinfo
  {journal} {Physical Review Letters}\ }\textbf {\bibinfo {volume} {98}},\
  \bibinfo {pages} {1} (\bibinfo {year} {2007})}\BibitemShut {NoStop}%
\bibitem [{\citenamefont {Ufimtsev}\ and\ \citenamefont
  {Martinez}(2009)}]{Ufimtsev2009/10.1021/ct9003004}%
  \BibitemOpen
  \bibfield  {author} {\bibinfo {author} {\bibfnamefont {I.~S.}\ \bibnamefont
  {Ufimtsev}}\ and\ \bibinfo {author} {\bibfnamefont {T.~J.}\ \bibnamefont
  {Martinez}},\ }\href {\doibase 10.1021/ct9003004} {\bibfield  {journal}
  {\bibinfo  {journal} {Journal of Chemical Theory and Computation}\ }\textbf
  {\bibinfo {volume} {5}},\ \bibinfo {pages} {2619} (\bibinfo {year}
  {2009})}\BibitemShut {NoStop}%
\bibitem [{\citenamefont {Gaus}, \citenamefont {Cui},\ and\ \citenamefont
  {Elstner}(2011)}]{Gaus2011/10.1021/ct100684s}%
  \BibitemOpen
  \bibfield  {author} {\bibinfo {author} {\bibfnamefont {M.}~\bibnamefont
  {Gaus}}, \bibinfo {author} {\bibfnamefont {Q.}~\bibnamefont {Cui}}, \ and\
  \bibinfo {author} {\bibfnamefont {M.}~\bibnamefont {Elstner}},\ }\href
  {\doibase 10.1021/ct100684s} {\bibfield  {journal} {\bibinfo  {journal}
  {Journal of Chemical Theory and Computation}\ }\textbf {\bibinfo {volume}
  {7}},\ \bibinfo {pages} {931} (\bibinfo {year} {2011})},\ \Eprint
  {http://arxiv.org/abs/NIHMS150003} {arXiv:NIHMS150003} \BibitemShut {NoStop}%
\bibitem [{\citenamefont {Aradi}, \citenamefont {Hourahine},\ and\
  \citenamefont {Frauenheim}(2007)}]{Aradi2007/10.1021/jp070186p}%
  \BibitemOpen
  \bibfield  {author} {\bibinfo {author} {\bibfnamefont {B.}~\bibnamefont
  {Aradi}}, \bibinfo {author} {\bibfnamefont {B.}~\bibnamefont {Hourahine}}, \
  and\ \bibinfo {author} {\bibfnamefont {T.}~\bibnamefont {Frauenheim}},\
  }\href {\doibase 10.1021/jp070186p} {\bibfield  {journal} {\bibinfo
  {journal} {Journal of Physical Chemistry A}\ }\textbf {\bibinfo {volume}
  {111}},\ \bibinfo {pages} {5678} (\bibinfo {year} {2007})}\BibitemShut
  {NoStop}%
\bibitem [{\citenamefont {Gaus}, \citenamefont {Goez},\ and\ \citenamefont
  {Elstner}(2013)}]{Gaus2013/10.1021/ct300849w}%
  \BibitemOpen
  \bibfield  {author} {\bibinfo {author} {\bibfnamefont {M.}~\bibnamefont
  {Gaus}}, \bibinfo {author} {\bibfnamefont {A.}~\bibnamefont {Goez}}, \ and\
  \bibinfo {author} {\bibfnamefont {M.}~\bibnamefont {Elstner}},\ }\href
  {\doibase 10.1021/ct300849w} {\bibfield  {journal} {\bibinfo  {journal}
  {Journal of Chemical Theory and Computation}\ }\textbf {\bibinfo {volume}
  {9}},\ \bibinfo {pages} {338} (\bibinfo {year} {2013})}\BibitemShut {NoStop}%
\bibitem [{\citenamefont {Jahangiri}\ \emph {et~al.}(2013)\citenamefont
  {Jahangiri}, \citenamefont {Dolgonos}, \citenamefont {Frauenheim},\ and\
  \citenamefont {Peslherbe}}]{Jahangir2013/10.1021/ct300919h}%
  \BibitemOpen
  \bibfield  {author} {\bibinfo {author} {\bibfnamefont {S.}~\bibnamefont
  {Jahangiri}}, \bibinfo {author} {\bibfnamefont {G.}~\bibnamefont {Dolgonos}},
  \bibinfo {author} {\bibfnamefont {T.}~\bibnamefont {Frauenheim}}, \ and\
  \bibinfo {author} {\bibfnamefont {G.~H.}\ \bibnamefont {Peslherbe}},\ }\href
  {\doibase 10.1021/ct300919h} {\bibfield  {journal} {\bibinfo  {journal}
  {Journal of Chemical Theory and Computation}\ }\textbf {\bibinfo {volume}
  {9}},\ \bibinfo {pages} {3321} (\bibinfo {year} {2013})}\BibitemShut
  {NoStop}%
\bibitem [{\citenamefont {Zhechkov}\ \emph {et~al.}(2005)\citenamefont
  {Zhechkov}, \citenamefont {Heine}, \citenamefont {Patchkovskii},
  \citenamefont {Seifert},\ and\ \citenamefont
  {Duarte}}]{Zhechkov2005/10.1021/ct050065y}%
  \BibitemOpen
  \bibfield  {author} {\bibinfo {author} {\bibfnamefont {L.}~\bibnamefont
  {Zhechkov}}, \bibinfo {author} {\bibfnamefont {T.}~\bibnamefont {Heine}},
  \bibinfo {author} {\bibfnamefont {S.}~\bibnamefont {Patchkovskii}}, \bibinfo
  {author} {\bibfnamefont {G.}~\bibnamefont {Seifert}}, \ and\ \bibinfo
  {author} {\bibfnamefont {H.~A.}\ \bibnamefont {Duarte}},\ }\href {\doibase
  10.1021/ct050065y} {\bibfield  {journal} {\bibinfo  {journal} {Journal of
  Chemical Theory and Computation}\ }\textbf {\bibinfo {volume} {1}},\ \bibinfo
  {pages} {841} (\bibinfo {year} {2005})}\BibitemShut {NoStop}%
\bibitem [{\citenamefont {Rapp{\'{e}}}\ \emph {et~al.}(1992)\citenamefont
  {Rapp{\'{e}}}, \citenamefont {Casewit}, \citenamefont {Colwell},
  \citenamefont {{Goddard III}},\ and\ \citenamefont
  {Skiff}}]{Rappe1992/10.1021/ja00051a040}%
  \BibitemOpen
  \bibfield  {author} {\bibinfo {author} {\bibfnamefont {A.~K.}\ \bibnamefont
  {Rapp{\'{e}}}}, \bibinfo {author} {\bibfnamefont {C.~J.}\ \bibnamefont
  {Casewit}}, \bibinfo {author} {\bibfnamefont {K.~S.}\ \bibnamefont
  {Colwell}}, \bibinfo {author} {\bibfnamefont {W.}~\bibnamefont {{Goddard
  III}}}, \ and\ \bibinfo {author} {\bibfnamefont {W.}~\bibnamefont {Skiff}},\
  }\href {\doibase 10.1021/ja00051a040} {\bibfield  {journal} {\bibinfo
  {journal} {J. Am. Chem. Soc.}\ }\textbf {\bibinfo {volume} {114}},\ \bibinfo
  {pages} {10024} (\bibinfo {year} {1992})}\BibitemShut {NoStop}%
\end{thebibliography}
\end{document}